\documentstyle[11pt,aaspp4]{article}

\def\xx{\enspace\enspace}

\def\et{et al.}
\def\kms{km s$^{-1}$}
\def\ha{H$\alpha$}

\def\solar{\ifmmode_{\mathord\odot}\;\else$_{\mathord\odot}\;$\fi}

\def\r25{R$_{25}$}
\def\coldens{atoms cm$^{-2}$}
\def\sigdens{M\solar pc$^{-2}$}
\def\sigcrit{$\Sigma_{c}$}
\def\siggas{$\Sigma_g$}
\def\sighii{$\Sigma_g$(max)}
\def\sighi{$\Sigma_{HI}$}
\def\sigha{$\Sigma_{H\alpha}$}
\def\sigrat{$\Sigma_g/\Sigma_{c}$}

\def\mhilb{M$_{HI}$/L$_B$}
\def\mhilh{M$_{HI}$/L$_{H(-0.5)}$}
\def\mhilbunit{M\solar/L$_{B,}$\solar}
\def\HII{H$\,${\sc ii}}

\def\sfrunit{M\solar\ yr$^{-1}$ kpc$^{-2}$}
\def\ergsec{erg s$^{-1}$}

\begin{document}

\title{Neutral Hydrogen and Star Formation in the Irregular Galaxy NGC 2366
}

\author{Deidre A.\ Hunter}
\affil{Lowell Observatory, 1400 West Mars Hill Road, Flagstaff, Arizona 86001
USA; \\dah@lowell.edu}

\author{Bruce G.\ Elmegreen}
\affil{IBM T.\ J.\ Watson Research Center, PO Box 218, Yorktown Heights,
New York 10598 USA; \\bge@watson.ibm.com}

\and

\author{Hugo van Woerden}
\affil{Kapteyn Astronomical Institute, Postbus 800, 9700 AV Groningen,
The Netherlands; hugo@astro.rug.nl}

\begin{abstract}

We present deep UBVJHK\ha\ images and HI maps of the 
irregular galaxy NGC 2366.
Optically, NGC 2366 is a boxy-shaped exponential disk
seen at high inclination angle.
The scale length and central surface brightness
of the disk are normal for late-type galaxies.
Although NGC 2366 has been classified as a barred Im galaxy, we do not see any
unambiguous observational signature of a bar.
There is an asymmetrical extension of stars along one end of the major axis
of the galaxy, and this is where the furthest star-forming
regions are found, at a radius of 1.3 times the Holmberg radius.
The star formation activity of the galaxy is dominated by the supergiant
\HII\ complex NGC 2363, but the 
global star formation rate for NGC 2366 is only moderately elevated
relative to other Im galaxies.
The star formation activity drops off with radius
approximately as the starlight in the inner part of the galaxy
but it drops faster in the outer part.

There are some peculiar features of the HI distribution and kinematics.
First, the integrated HI shows two ridges running parallel to the major axis
that when deprojected appear as a large ring. 
Second, the
velocity field exhibits several large-scale anomalies
superposed on a rotating disk;
some of these may be from a weak bar
that has no Inner Lindblad Resonance.
Third, the inclination and position angles derived from the kinematics differ
from those derived from the optical and HI morphology.
Fourth, there 
are regions in the HI of unusually high velocity dispersion, but these
regions are not associated with the optical galaxy nor any obvious HI feature.
Instead the velocity dispersion
correlates with a deficit of HI emission in a manner suggestive of
long-range, turbulent pressure equilibrium.
In other respects the HI is fairly normal. 
The azimuthally-averaged surface density of HI is comparable to that of other
irregulars in the inner part of the galaxy but drops off slower and
extends further in the outer parts.
The HI around the star-forming complex NGC 2363 is fairly unremarkable.

Like in other disk galaxies, the gas in NGC 2366 is lumpy and star-forming regions
are associated with these HI complexes. \HII\ regions are found where the gas 
densities locally exceed 6 \sigdens. 
This threshold is required to provide a cool phase of HI as a first
step toward star formation. 
NGC 2366, like other irregulars, has low gas densities relative
to the critical gas densities of gravitational instability models,
so large-scale gravitational instabilities operate slowly or not
at all.  Considering the lack of shear in the optical part of this
galaxy, the relative slowness of such instabilities may not be a problem
-- there is little competition to 
the slow gravitational contraction that follows energy dissipation.  
This differs from the situation in giant spiral disks where
the shear time is short, comparable to the energy dissipation time, and
strong self-gravity is required for a condensation to 
grow and dissipate its turbulent energy before it shears away.

The sub-threshold surface densities are also not unusual if they
are viewed using the critical tidal density for
gravitational self-binding of a rotating cloud, rather than the critical
surface density from the usual disk instability condition.  The
peak densities in all regions of star formation are equal to the local
tidal densities, giving an agreement between these two quantities that
is much better than between the surface density and the critical value.
Evidently the large scale gas concentrations are all marginally bound
against background galactic tidal forces.  This condition for self-binding
may be more fundamental than the instability condition because it
is local, three-dimensional, and does not involve spiral arm generation
as an intermediate step toward star formation.

\end{abstract}


\section{Introduction}

NGC 2366 is classed as a barred Magellanic-type irregular (IBm) 
galaxy (de Vaucouleurs \et\ 1991).
The distance of 3.44 Mpc that we adopt here,
determined by Tolstoy \et\ (1995) from cepheid light curves,
places NGC 2366 in the M81 Group of galaxies
(Karachentseva, Karachentsev, \& B\"orngen 1985).
The nearest galaxy is the Scd NGC 2403 at 290 kpc.
NGC 2366 has an M$_B$ of $-$16.5 which is close to that
of the Small Magellanic Cloud, and
in terms of color, luminosity, and size NGC 2366 is a typical irregular
galaxy (Hunter 1997).
However, NGC 2366
is unusual in that it contains the supergiant \HII\ complex NGC 2363
at the southwest end of the galaxy (Youngblood \& Hunter 1999).
NGC 2363
has the same brightness in \ha, corrected for extinction, 
as the prototype supergiant
\HII\ complex, the 30 Doradus nebula in the
Large Magellanic Cloud, and contains a large fraction of the total
current star-formation activity of the galaxy (Aparicio \et\ 1995,
Drissen \et\ 2000).
However, there is also another very 
large \HII\ complex to the west of NGC 2363 as
well as numerous smaller \HII\ regions scattered along
the disk. Because of the presence of these two very large \HII\ complexes,
NGC 2366 has an inferred star formation rate that is towards the high
end of the distribution for normal Im galaxies.

In this paper we present optical, near-infrared, and HI radio interferometer
images of NGC 2366 for a comprehensive view of the galaxy.
We analyze them to learn about
the star formation processes in this galaxy.

\section{The Observations}

\subsection{The HI}

We obtained 21 cm line-emission observations of NGC 2366 with the 
Very Large Array 
(VLA\footnote{
The VLA is a facility of the National Radio Astronomy Observatory (NRAO),
itself a facility of the National Science Foundation that is operated
by Associated Universities, Inc.})
radio interferometer
in its C and D-array configurations.
The C-array observations were made
1989 July 21 with a total on-source time of 235 minutes.
The D-array observations were made 1989 November 7 with
a total on-source time of 331 minutes. The total bandwidth was
3.125 MHz with a channel separation of 24.414 kHz, that equals 5.16 \kms.
However, a total of only 64 channels was allowed at that time
because of computer
limitations, and so
the velocity coverage is only 330 \kms.
The data were on-line Hanning-smoothed, and the resulting velocity
resolution is 10 \kms.
The central heliocentric velocity was 107 \kms.
The field center is 7$^h$ 23$^m$ 34.2$^s$, $+$69\arcdeg\
18\arcmin\ 42\arcsec\ (equinox 1950, here and throughout).
 
We subtracted the continuum emission in the {\it uv}-plane using
line-free channels on either end of the spectrum and combined
the data from the different arrays.
To make maps, we employed a routine in NRAO's
Astronomical Image Processing System (AIPS)
that enables one to choose a sample weighting that is
in between the standard ``natural'' weighting, which gives the
highest signal-to-noise but at the expense of long wings in the
beam, and ``uniform'' weighting, which gives the best resolution
but at the cost of signal-to-noise and the presence of negative
sidelobes. 
We chose a sample weighting
that gives a formal increase
of only 5\% in noise yet with a significant improvement in beam
profile over what would have been achieved with
``natural'' weighting.  The resulting synthesized beam profile
(FWHM) is 33.6\arcsec$\times$28.9\arcsec. This corresponds to
560 pc$\times$480 pc at the galaxy.
We deconvolved the
maps until there were roughly comparable numbers of positive and
negative components.
The rms noise in the cleaned channel maps is 1.09 mJy beam$^{-1}$.
We also constructed a map cube from the C-array data by itself in order
to take advantage of that array's higher resolution. The resulting
synthesized beam profile of those maps is
17.8\arcsec$\times$13.2\arcsec\ (300 pc$\times$220 pc at the galaxy), 
and the rms noise in a cleaned
channel map is 1.9 mJy beam$^{-1}$.
(Channel maps of the C$+$D-array HI data cube are 
discussed in Section \S \ref{sechi}).

To remove the portions of each channel map without emission that
are only contributing noise to the map, we used
the maps smoothed to twice the beamsize for a conditional transfer of
data in the unsmoothed maps. The channel maps were blanked wherever
the flux in the smoothed maps fell below 2.5$\sigma$.
Flux-weighted moment maps were made from the resulting data cube.
 
The flux in each channel of the C$+$D-array 
data cube was integrated over
a square 30\arcmin\ on a side, corresponding approximately to the diameter
of the primary beam of the telescope, in order to determine the total
HI flux detected in the maps.
The fluxes were corrected
for attenuation by the primary beam.
To compare with a single-dish observation we integrated the flux
over a square 10\arcmin\ on a side as well.
These integrated profiles are shown and compared to the single-dish
observation of Hunter, Gallagher, \& Rautenkranz (1982) 
in Figure~\ref{figsingle}.
Our integrated flux from the 10\arcmin\ box 
has a peak that is 12\% higher than the peak from the single-dish
telescope with a 10\arcmin\ beam.
The total integrated flux profile (over 30\arcmin) 
is about 36\% higher than the
single-dish profile.
Therefore, we recover all of the gas observed in the 10\arcmin\
single-dish observation as well as more gas extended beyond that.

The total HI mass detected within the primary beam of the VLA
observations is 8.0$\times10^8$ M\solar. This is smaller by a factor
of 10 than what was measured by Thuan \& Martin (1981) using the
NRAO 91 m telescope (10\arcmin\ beam); only a few percent
different from the mass reported by Huchtmeier, Seiradakis,
\& Materne (1981)
from single-dish mapping;
1.5 times larger than what is reported by Wevers, van der Kruit,
\& Allen (1986) 
from single-dish mapping;
and 10\% larger than the mass given by Swaters (1999) 
from single-dish mapping.
Thus, our integrated HI mass is close to values determined
by others; the Thuan and Martin value appears to be mistaken.

\subsection{Optical Images}

Deep UBV images were kindly obtained for us by P.\ Massey using
a Tektronix 2048$\times$2048 CCD on the Kitt Peak National Observatory
4 m telescope 1998 January. 
The telescope position was offset 30\arcsec\ between
the three images taken in each filter in order to improve the 
final flat-fielding.
Exposure times of 1800 s, 480 s, and 180 s for UBV, respectively, were
used.  The electronic pedestal was subtracted using the overscan strip,
and the images were flat-fielded using dome and sky flats.
Landolt (1992) standard stars were used to calibrate the photometry.
The pixel scale was 0.42\arcsec, the seeing was $\sim$1.6\arcsec.
 
Deep \ha\ images of NGC 2366 were obtained using the Perkins 1.8 m telescope
at Lowell Observatory during 3 observing runs 1994 to 1995. The observations
used an 800$\times$800 TI CCD provided to Lowell Observatory by the
National Science Foundation, the Ohio State University Fabry-Perot that
was used simply as a 3:1 focal reducer,
an \ha\ filter with a FWHM of 32 Angstroms, and an off-band filter centered
at 6440 Angstroms with a FWHM of 95 Angstroms. 
The off-band filter was used to subtract
stellar continuum from the \ha\ filter to leave only \ha\ nebular emission.
In order to survey the entire galaxy, 6 separate pointings were mosaicked
to produce the final \ha\ image. Observations consisted of several
1800 s or 3000 s images per pointing. The pixel scale was 0.49\arcsec,
and the seeing was $\sim$2.3\arcsec.

\subsection{Near-Infrared Images}

Near-infrared JHK images of NGC 2366 were obtained using the Ohio State
Infrared Imager-Spectrograph on the Perkins 1.8 m telescope
in two observing runs in 1996 and 1997. The pixel scale was 1.5\arcsec,
the field of view was 6.4\arcmin, and the effective seeing was 3\arcsec.
Observations alternated between the galaxy and a separate sky
position every 1--2 minutes. Upon each return to the galaxy, the position was
offset by 10\arcsec\ around the chip so that the numerous bad
pixels would not appear in the same place each time and to aid in removal
of stars in the sky frame. A correction for deviation from linearity
with increasing signal level
was applied to each of the 4 detectors,
observations of a dark filter were subtracted,
and flat-field frames were used to remove pixel-to-pixel variations.
The flat-field was determined by observations of a white screen in the dome,
and images made with the illuminating lamp off were subtraced from the images
taken with the lamp on.
Sky-subtracted
images were aligned and combined. The ``UKIRT'' standard stars
from the list compiled by S.\ Courteau were used
to calibrate the photometry, and so the photometry is on the
UKIRT system. The California Institute of Technology
(CIT) photometric system is 8\% bluer in J$-$H
and 4\% bluer in H$-$K compared to the UKIRT system.

A check on our photometry is provided by comparison to aperture 
photometry done by others. 
Aaronson, Mould, \& Huchra (1980) measured an H magnitude
of 10.44$\pm$0.08 for NGC 2366 through a 165.2\arcsec\ diameter aperture.
We measure an H magnitude that is 0.37 magnitudes fainter.
Their measurement through an aperture of 110.6\arcsec\ diameter
is also brighter.
On the other hand, our H-band measurement integrated to a radius of 117.5\arcsec\
differs by only 0.01 magnitude from that given by Pierini \et\ 
(1997). 

\section{Radio Continuum}

To examine continuum objects in the field of view of the radio maps,
we constructed and deconvolved maps from the C$+$D-array
{\it uv}-data before the
continuum was subtracted.
We summed channels not containing HI emission from NGC 2366.
The 21 cm continuum emission within the primary beam of the VLA 
is shown in Figure \ref{figcontfull}, 
and the fluxes of
all of the sources in the field of view of the primary beam
are given in Table \ref{tabcont}.
Only source 7 lies within the field of view of our V-band image,
and we do not detect any corresponding optical object there.

In Figure \ref{figcont} we show contours of the continuum emission
superposed
on the \ha\ image of the galaxy, 
In NGC 2366, we have primarily detected the supergiant \HII\ region
NGC 2363 although there is some continuum emission in the direction
of the giant \HII\ region to the west of NGC 2363 and some emission
extending to the north of NGC 2363.
Radio continuum observations of NGC 2363 at other wavelengths have 
shown the emission to be that of an optically thin, ionized gas
(Klein \& Gr\"ave 1986), as expected.

\section{The Optical and Near-Infrared View of the Galaxy}

\subsection{General Morphology}

Our V-band image of NGC 2366 is shown in Figures \ref{figouterbar} and
\ref{figallbars}. In the deep display (Figure \ref{figouterbar})
one can see that NGC 2366
is elliptical-shaped with a high degree of ellipticity implying a
relatively high inclination angle.  
Although the bulk of the galaxy is contained within regular isophotes,
there is an extension of starlight at a faint surface brightness level 
along the major axis to the southwest.  
This extension is highlighted by a superposed contour at 
26.54 magnitudes arcsec$^{-2}$ in
Figure \ref{figouterbar}.
This low surface brightness starlight is not mirrored
at the northeast end of the major axis.
The shallower display of the V-band image (Figure \ref{figallbars})
shows an inner cigar-shaped
morphology dominated by the giant star-forming region NGC 2363 at the
southwest end of the galaxy.
An intermediate surface brightness contour in Figure \ref{figallbars}
(the second faintest one)
illustrates an additional asymmetry of starlight within the galaxy
that skews the contour to the northwest in the northeastern part of
the galaxy and to the southeast in the southwestern part of the galaxy.
This skewing could be described as an asymmetrical warp, but it is
not found in the outer contours.

\subsection{Bar Structures}

NGC 2366 is classified as a barred Im galaxy and the cigar-shaped 
morphology contributes to the impression that the galaxy is barred.
However, identifying bar structures in irregular galaxies is difficult
from morphology alone.
As discussed by Roye \& Hunter (2000), this identification 
is complicated by the lack 
of symmetry provided by
spiral arms coming out of the nucleus or bar. It is particularly
complicated in galaxies that are also inclined and often boxy in shape.
This is especially difficult if the bar is comparable in size
to the entire galaxy as appears to be the case in, for example, NGC 4449
(Hunter, van Woerden, \& Gallagher 1999).
Does NGC 2366 contain an over-sized bar structure, 
or does it just appear bar-like
because it is highly inclined?

We have explored the possible presence of bars in NGC 2366 by
fitting elliptical structures to isophotes.
According to Athanassoula \et\ (1990), a bar can be fit
with a curve of the form
$({{|x|}\over{a}})^c + ({{|y|}\over{b}})^c = 1$,
where $a$ and $b$ are the semi-major and minor axes
and $c$ is a parameter that describes the boxiness of the
bar structure. For $c=2$ the structure is an ellipse,
for $c<2$ the structure is ``disky,'' and for
$c>2$ the structure is ``boxy.'' 
In NGC 2366, we have fit the inner and outer parts well with a curve with
$c=3.0$, indicating a boxy shape.

First, we fit the outer isophote at 25.66 magnitudes arcsec$^{-2}$
(seen as the inner of the two isophotes in Figure \ref{figouterbar})
with a curve that is shown
superposed on the V-band image in Figure \ref{figouterbar}
as a thick smooth contour. 
The isophotes in Figure \ref{figouterbar} are determined 
from a smoothed version of the V-band image to improve
signal-to-noise in the outer parts. 
One can see that, if we exclude the faint extension to the southwest
encompassed by the fainter isophote at 26.54 magnitudes arcsec$^{-2}$,
the contour at 25.66 magnitudes arcsec$^{-2}$
nicely encompasses the bulk of the optical galaxy.
This outer ``bar'' has a semi-major axis of 1.6R$_{25}$ at a position angle of 
32.5\arcdeg, where R$_{25}$ is the radius of the galaxy
at a surface brightness of 25
magnitude arcsec$^{-2}$ in B. The minor-to-major axis ratio $b/a$ is 0.42. 
If the $b/a$ ratio reflects the inclination of the galaxy
and the intrinsic ratio of the disk is taken to be 0.3
(Hodge \& Hitchcock 1966, van den Bergh 1988),
the inclination shown by
the outer galaxy is 72\arcdeg.

Second, we fit bar structures to two inner isophotes 
that are more normal in size compared to bars in spirals.
These are shown as thick contours in the inner part of
Figure \ref{figallbars}. 
The thick outer contour is reproduced from Figure \ref{figouterbar}.
One of the inner fits excludes
the supergiant \HII\ region NGC 2363 and the other includes it
since it has been argued that the formation of NGC 2363 may have
been facilitated by placement at the end of a bar
(Elmegreen \& Elmegreen 1980).
The sizes, position angles, $b/a$, and centers of these two bar
fits, as well as characteristics of the fit 
to the outer galaxy, are listed in Table \ref{tabpa}.
One can see that all three fits have nearly the same
position angle. We are not seeing a twisting of the isophotes
from the inner to the outer galaxy which in spiral galaxies is a good
clue to the presence of a bar. 
The center of the middle isophote is offset from the center of the
outer isophote 19\arcsec\ east and 8\arcsec\ north. The center of the
inner
isophote is offset 33\arcsec\ east and 31\arcsec\ north
because the inner one does not include NGC 2363.

Figure \ref{figallbars} also shows the observed isophotes at 
26.54, 25.66, 24.04, 23.04, and 22.68 magnitudes arcsec$^{-2}$.
The contour at 24.04 mag arcsec$^{-2}$  illustrates
the skewing of the V-band image that was
discussed above. We will return to this skewed contour
in section \ref{sect:revisit} below.

In addition to these three carefully selected isophotes that we fit
with ellipses by hand, we have done automatic elliptical isophote fitting
of the entire V-band image of the galaxy. We used annuli of 63\arcsec\
width out to a radius of 7.4\arcmin, 
and allowed the fitting procedure to fit position, ellipticity,
and position angle for each ellipse. This routine does not allow
the boxy structures that we used above---only ellipses. The results
of this type of general
fitting procedure should be viewed with caution in lumpy galaxies
like NGC 2366, where young regions here and there can highly influence
the fitting. We found that the inclination angle varied from
68\arcdeg\ to 79\arcdeg\ and the position angle varied from 29\arcdeg\
to 34\arcdeg. Most of the variation---a decrease outward 
in position angle
and inclination---occurred in the inner 3\arcmin\
of the galaxy. Some of this variation is undoubtedly due to the
difficulty in fitting just mentioned, but it is also possible that
some of the variation 
is the signature of a bar structure with a semi-major axis
of $\sim$3\arcmin.

If the outer isophote is a true bar, its $b/a$ ratio of 0.42 falls within
the range of what is observed for bars in late-type spirals (Martin 1995).
However, $a$/R$_{25}$ is 1.6 in NGC 2366.
In spirals the ratio $a$/R$_{25}$ is much smaller:
0.2 for Sd--Sm galaxies (Elmegreen \et\ 1996),
0.05--0.25 for Sd--Sdm spirals (Martin 1995),
and 0.25--0.3 for Sd's with exponential bars (Elmegreen \& Elmegreen 1985).
The bar fit to the inner isophote would be reasonable in size
compared to these values, while the middle isophote fit would
still be a bit large by comparison.

Surface brightness cuts, 33\arcsec\ wide,
along the major and minor axis of the inner bar
are shown in Figure \ref{figbarcut}. 
Neither cut can be reasonably fit with a Gaussian profile,
as they are in a sample of early-type spirals observed by Ohta,
Hamabe, \& Wakamatsu (1990).
The shapes of the minor and major axis cuts individually resemble a
few of those observed in late-type barred spirals by Elmegreen \et\ (1996).
Among barred late-type spirals (Combes \& Elmegreen 1993,
Elmegreen \et\ 1996) the bar is commonly well-fit with an
exponential whereas in early-type spirals the bars tend to be more
uniform in intensity.

Thus, we find three isophotes that could be claimed to be bar
structures: an inner one that places the supergiant \HII\ region NGC 2363
at one end, a middle structure that includes NGC 2363, and an outer structure
that encompasses most of the galaxy. From the surface photometry there is no
compelling evidence that any of these structures is a true bar, but
we will return to this issue below when we present the HI data and the
velocity field of the gas.

\subsection{Surface Photometry}

UBVJHK surface photometry is shown in Figures \ref{figubv} and 
\ref{figjhk} (see also Wevers \et\ 1986). 
We used ellipses with a position angle of 32.5\arcdeg,
ellipticity corresponding to an inclination of 72\arcdeg, and
a semi-major axis step size of 21\arcsec.
The inclination, position angle, and center were determined from the outer
isophote shown in Figure \ref{figouterbar}. 
The surface photometry and colors have been corrected for reddening using
a total E(B$-$V)$_t$=E(B$-$V)$_f$$+$0.05, where the foreground reddening
E(B$-$V)$_f$ is 0.043 (Burstein \& Heiles 1984). We use the reddening
law of Cardelli, Clayton, \& Mathis (1989) and A$_V$/E(B$-$V)$=$3.1.
Thus, A$_V$ is 0.29 and E(U$-$B) is 0.06.

We determined 
R$_{25}$ to be 2.6\arcmin\ ($=$2.6 kpc).
Our radius is 27\% bigger than that given by de Vaucouleurs \et\ (1991).
The Holmberg radius, R$_H$, originally defined to a photographic surface
brightness, is measured at an equivalent B-band surface brightness
$\mu_B = 26.7 - 0.149(B-V)$. For a (B$-$V)$_0$ of 0.27, the
Holmberg radius is determined at a $\mu_B^0$ of 26.66 magnitudes
arcsec$^{-2}$. We find that R$_H$ is 5.1\arcmin\ ($=$5.1 kpc). Both
R$_{25}$ and R$_H$ are determined from the reddening corrected
surface photometry. Global properties of NGC 2366 
are listed in Table \ref{tabglobal}.

We fit an exponential disk to $\mu_V^0$, excluding the central 
1.2\arcmin, and the fit is shown in 
Figure \ref{figubv}. 
Except for the central region, NGC 2366 is fit well with an exponential disk profile having 
a central surface brightness
$\mu_0$ of 22.82$\pm$0.09 magnitudes arcsec$^{-2}$
and a disk scale length R$_D$ of 1.59$\pm$0.04 kpc.
This same exponential also fits $\mu_J$ reasonably well
over the radii where the J-band is detected.
Noeske \et\ (2000) have found a scale length of  half this---
0.88$\pm$0.03 kpc and 0.82$\pm$0.03 kpc for B and I-bands, respectively---
from fitting the inner 2.7\arcmin\ of the galaxy.
De Jong (1996) has determined disk characteristics of a large sample
of galaxies spanning a wide range in type.
NGC 2366 has a central surface brightness, corrected to the B passband,
and an R$_D$ that are in the middle of the range of what de Jong finds for
late-type galaxies.

The integrated colors of NGC 2366 are similar to 
those of the luminous irregular
NGC 4449 with the exception that NGC 2366 is 0.1--0.2
magnitude bluer in the near-infrared (Hunter \et\ 1999).
Like in most irregulars, colors in NGC 2366 are fairly constant with radius
(see also Wevers 1984).
Integrated values are given in Table \ref{tabglobal}, and annular
colors are shown in Figures \ref{figubv} and \ref{figjhk}.

These figures indicate that NGC 2366 does not exhibit any
major color gradient with radius.
It is common lore
that irregular galaxies do not have color gradients, and NGC 2366 
appears to be perfectly consistent with that, now frequent, assumption.
However, in the case of NGC 2366, and by extension other irregulars,
azimuthally-averaged colors do not in fact tell the whole story.
In Figure \ref{figdivbv} we show a B/V ratio image, averaged $3\times3$
pixels to improve the signal-to-noise.
Although the galaxy is relatively constant in color with radius in an
average sense, there are color variations on large scales. 
(The red circular region in the Figure at
$\sim$7$^h$ 23$^m$ 30$^s$, 69\arcdeg\ 21\arcmin\ is due to bright
foreground stars.)
In Figure \ref{figdivbv} one can see a long, blue ridge 
running the extent
of the galaxy, including the central bars identified in Figure \ref{figallbars}
and the unusual, off-axis higher surface brightness extensions to the
northwest and southeast.
A 1.5 kpc diameter region $\sim$50\arcsec\ to the west/southwest of 
the center
has a redder stellar population.
This coincides with the center of the HI ring just to the north of
NGC 2363 discussed below.
It is 0.1--0.15 mag redder than the bluer areas of the galaxy
outside of \HII\ regions.
Another large, low surface brightness region $\sim2.5$\arcmin\
to the northeast of center
also has a red stellar population. 
The western part of NGC 2363 also stands out as red, presumably
due to the dust that obscures the massive star cluster there (Drissen \et\
2000).
The faint southwest extension of stars outlined by the outer isophote
in Figure \ref{figouterbar}
has colors like those of
most of the rest of the background galaxy that does not contain \HII\ regions.

\subsection{Star-Forming Regions}

Our \ha\ image of NGC 2366 is shown in Figure \ref{figcont},
and the central portion was also shown by Hunter, Hawley,
\& Gallagher (1993) in their study of the extra-\HII\ region
ionized gas. Although there are numerous star-forming regions
in NGC 2366,
the \ha\ morphology is dominated by two very large \HII\ complexes:
NGC 2363 (numbered I and II by Drissen \et\ 2000) and a second one directly to 
the west of NGC 2363 (numbered III by Drissen \et\ 2000).
Supergiant \HII\ regions like these are rare among normal irregular
galaxies (Youngblood \& Hunter 1999).

NGC 2363 has an L$_{H\alpha,0}$ of 1.3$\times10^{40}$ ergs s$^{-1}$.
This represents 70\% of the galaxy's total \ha\ luminosity.
NGC 2363 has the \ha\ equivalent of the 30 Doradus nebula,
where 30 Doradus is the supergiant \HII\ region in the Large
Magellanic Cloud that is often taken as defining the class
of supergiant \HII\ complexes. (The extinction corrected \ha\ luminosity
of 30 Doradus is taken as $1.5\times10^{40}$ \ergsec, from Kennicutt 1984).
The \ha\ luminosity is what would be expected of 1260 O7V stars.

Figure \ref{figcont} does not include two
tiny \HII\ regions that are located south of NGC 2363
by 2.7\arcmin\ and 3.0\arcmin. (They can be seen in Figure
\ref{fighionha} near 7$^h$ 23$^m$, 69\arcdeg\ 14\arcmin).
They are in fact
the furthest \HII\ regions from the center of the galaxy
and are located at 6.6 kpc and 5.9 kpc radius in the plane of the
galaxy. The furthest \HII\ region corresponds to a distance
of 2.5R$_{25}$ or
1.3R$_H$. They are part of the faint extension of starlight
along the major axis to the southwest seen in the V-band in 
Figure \ref{figouterbar}.

In Figure \ref{figha} we show the azimuthally-averaged \ha\ surface
brightness and compare it to $\mu_V^0$.
Once past the \HII\ region NGC 2363, the \ha\ surface brightness first drops 
off at a rate roughly
comparable to the starlight, but then beyond 4.5 kpc drops off faster. The
current star formation activity ends before the detected starlight ends.
The outer 16\% of the galaxy by area in which we detect starlight
contains no detectable \HII\ region.

The \ha\ luminosity and derived star formation rates of the galaxy are
given in Table \ref{tabglobal}. NGC 2366 has a star formation
rate per unit area that is normal, but towards the high end of the 
range, for irregular galaxies (Hunter 1997).
At its current rate of consumption, the galaxy can turn gas
into stars for another 8 Gyr if all of the gas
associated with the galaxy can be used. The timescale to run
out of gas becomes 12--32 Gyr if recycling of gas from dying
stars is also considered (Kennicutt, Tamblyn, \& Congdon 1994). 
These timescales to exhaust the gas
are somewhat short for irregular galaxies
(Hunter 1997), but, nevertheless, the galaxy is in no danger of
running out of the fuel for star formation any time soon.

\section{The HI} \label{sechi}

NGC 2366 was first mapped in HI with the single-dish radio
telescope at Effelsberg (Huchtmeier \et\ 1981).
They reported that the HI covers 30\arcmin.
Wevers \et\ (1986) presented Westerbork
observations with a 25\arcsec\ beam and 16.5 \kms\ channel
separation, and recently,
Swaters (1999) has presented new Westerbork observations with
a 30\arcsec\ beam and 2--4 \kms\ channel separation. 
Braun (1995) discussed 9\arcsec\ resolution VLA B$+$C$+$D-array observations
with 5.16 \kms\ channel separation.
Our VLA C$+$D-array observations, with much longer integration times
than Braun's data, contribute medium spatial resolution
observations with fairly high velocity resolution at high sensitivity.

Channel maps of the C$+$D-array HI data cube are presented in 
Figure \ref{figchan}.
We detect HI emission from 24.5 \kms\ to 174.0 \kms, and one can
see general rotation from the northeast, at higher radial velocities,
to the southwest, at lower radial velocities.
There is also some interesting low column density emission that
appears in the northeast part of the galaxy at velocities 102--138 \kms.
There may be a weak counterpart to the south, at velocities 61--81 \kms.
One ridge of HI is clearly seen at velocities of 61--117 \kms,
and a second ridge at 117--153 \kms.

The integrated HI is shown in 
Figure \ref{figgrayfull}---two displays to show both the fainter emission in the
outer parts and the inner, higher column density material
near the center. To help the reader
orient the HI with respect to the
optical galaxy, Figure \ref{figgrayfull} shows 
one of the contours from the V-band
image (Figure \ref{figouterbar}).
We detect HI emission to a radius of 11.7\arcmin\ at a
column density of 1.7$\times10^{19}$ cm$^{-2}$.
Huchtmeier \et\ (1981) reported that the HI in NGC 2366 is extended
to a radius of 15\arcmin\ at
10$^{19}$ \coldens. 
However, in view of the smearing introduced by the Effelsberg
beam (FWHM$\sim$8.8\arcmin), the results could be consistent.
VLA maps of the extended HI around NGC 2366 are under analysis
and will settle this issue.

In the integrated HI map, one can see large clumps of gas within
the optical galaxy, arranged approximately in two parallel ridges of HI.
Deprojected, the HI appears as a ring with a 
depression in the middle, similar to Sextans A
(Skillman \et\ 1988). The ring has a diameter of 5.5 kpc.
Two-dimensional ratio images of the B and V-bands suggest that
the center of this ring is redder than the surrounding stellar populations
(Figure \ref{figdivbv}).
The center of the ring (7$^h$ 23$^m$ 31.3$^s$, 69\arcdeg\ 19\arcmin\
53\arcsec) is to the north of the kinematical center and
the center of the outer optical isophote by a little more than 1\arcmin.

The ridges of gas are surrounded by a more extended relatively smooth 
distribution of gas.
However, in the outer parts we do see a cloud in the southwest surrounded by 
more diffuse emission. In the northeast we see a somewhat unusual
extension that looks like a ``knob'' extending east from the galaxy with
some filamentary material extending south of it.
There is also some filamentary HI emission to the north of the galaxy
center around 7$^h$ 23$^m$ 45$^s$, 69\arcdeg\ 24\arcmin.

One can also see that the outer shape of the HI distribution is generally
consistent with that of the optical galaxy.
Specifically, the morphology of the integrated HI indicates a position angle of
23.5\arcdeg\ and $b/a$ of 0.41. This position angle is 9\arcdeg\
smaller (HI oriented further towards the north)
than what we see for the optical galaxy, but the 
minor-to-major axis ratio is the same.
However, the HI is not centered symmetrically on the optical galaxy.
The HI extends further to the northeast than to the southwest relative
to the V-band isophote:
The center of the HI morphology is shifted 47\arcsec\ to the east
and 54\arcsec\ to the north from the center of the outer optical
isophotes. 
However, asymmetric distributions of HI are common
among disk galaxies (Richter \& Sancisi 1994).

\subsection{Surface Density}

In Figure \ref{figsurden} we show the azimuthally-averaged surface density
of
the HI gas in NGC 2366. 
We have integrated the HI in 25\arcsec\ radial steps.
In determining the surface density, we had
to choose a position angle and
inclination which is somewhat problematical given the selection available
in Table \ref{tabpa}, particularly including the different 
values we will find below from the kinematics of the HI.
In this plot we show the surface density that results from two choices
of position angle and inclination: 1) the values that result from
the HI morphology (23\arcdeg\ position angle and 73\arcdeg\
inclination) and 2) the values that result from 
the kinematics of the HI (46\arcdeg\ position angle and
65\arcdeg\ inclination).
The choice of the kinematic parameters implies that the HI distribution
is not circularly symmetric but elongated.
The choice of the morphologically-derived parameters implies
that the HI distribution is axisymmetric, but that there are
non-circular motions. 
We present both cases, but 
the conclusions are the same for either choice of parameters.

From the Figure we see that, outside the center ($>$1.75\arcmin),
the gas surface density drops off fairly smoothly.
In order to compare the surface brightnesses in HI, \ha, and V-band,
we have plotted the logarithm of $\Sigma_{HI}$, the logarithm
of $\Sigma_{H\alpha,0}$, and $\mu_V^0$ in magnitudes over the
same logarithmic interval. These are shown in the right panel
of Figure \ref{figsurden}. There we can see that the HI surface
density drops off much more slowly than the starlight or star
formation activity. This is also what we found for the IBm galaxy
NGC 4449 (Hunter \et\ 1999).

In the left panel of 
Figure \ref{figsurden} we compare the gas surface density profile
of NGC 2366 to that of average Im, Sdm, and Sd galaxies.
The average profile for Im galaxies comes from Broeils \& van Woerden
(1994) and those for the spirals from Cayatte \et\ (1994) who obtained
the data for individual galaxies from Broeils and van Woerden
and Warmels (1988).
This type of comparison requires that the observations be at
similar physical spatial resolutions and that there be many resolution elements
across the galaxy. The beam of our NGC 2366 observations is small compared
to the HI extent of the galaxy which is about 21 beam-widths in radius.
The observations of Im galaxies by Broeils and
van Woerden have HPBWs of 340--1020 pc with a median of 400 pc
(for an H$_0$ of 65 \kms\ Mpc$^{-1}$). This is comparable to our NGC 2366
resolution of 
560 pc$\times$480.
The HPBWs of the observations of spirals, on the other hand, vary from 
530 pc to 5190 pc (median of 2600 pc for Sdm and 1070 pc for
Sd galaxies).
Therefore, the profiles for the spirals
are affected by a larger beam size compared to that of NGC 2366.

What we see in Figure \ref{figsurden}
is that the HI surface density of NGC 2366 is comparable
to the average of a sample of Im galaxies in the inner 3.5\arcmin.
The NGC 2363 complex is dominating at
$R/R_{25}\sim0.7$.
Beyond 3.5\arcmin\ the HI falls off more slowly than in other Im galaxies and
the HI extends much further.
The spiral samples on average have much
lower surface densities in the center, perhaps because of resolution
effects and the fact that the gas is dominated by molecular material
rather than atomic (Honma, Sofue, \& Arimoto 1995), 
and drop quickly at smaller radii.

\subsection{Relationship of HI to Integrated Starlight}

In NGC 2366 the ratio of integrated HI content to starlight, \mhilb,
is 1.36 \mhilbunit.
The average \mhilb\ ratio measured by Broeils \& Rhee (1997) from their
sample of galaxies 
is 0.36$\pm$0.47 \mhilbunit\ for all morphological types
and 0.32 for Im galaxies.
For a large sample of irregulars Roberts \& Haynes (1994; see also Roberts
1969)
show a median \mhilb\
of about 0.6 \mhilbunit\ for Im galaxies.
NGC 2366, therefore, appears to be unusually gas rich for its luminosity
by a factor of $\sim$2--4.

Bothun (1984) has suggested that a better ratio to consider is \mhilh\
where L$_{H(-0.5)}$ is the luminosity of the galaxy in the H-band within 
an aperture $A$ of $\log A/D=-0.5$, where $D$ is the size of the galaxy
in the optical. As originally defined, $A$ corresponds to a radius
of 0.316R$_0$ where R$_0$ is close to \r25.
Bothun showed that \mhilh\ is a function of the luminosity of the galaxy.
For a galaxy like NGC 2366 with a total M$_{B,0}$ of $-$17, the ratio \mhilh\
would be expected to be of order 1.8, and
Broeils \& Rhee (1997) measured an average \mhilh\ of
0.5$\pm$0.5 for their sample of galaxies.
Instead, we measure the \mhilh\ ratio to be 13, 
which is 7 times higher than expected for this luminosity.
In this comparison NGC 2366 is very much more gas-rich
than other galaxies for its luminosity.
The integrated HI measurements of spirals used by Broeils and Rhee from 
interferometer data agree within a few percent with single-dish observations
(Broeils \& van Woerden 1994) and thus it is unlikely that they are missing
significant amounts of HI. On the other hand, their Figure 1 shows 
\mhilh\ increasing with Hubble type and they do not include galaxies later
than type Sd, for which the average \mhilh\ is $\sim$1. Thus, it could
be that NGC 2366 is not too unusual for its Hubble type, but it is certainly
much gassier than Sd-type galaxies. More H-band observations of irregulars
are needed to determine how unusual are NGC 2366 and NGC 4449, also with
an unusually high \mhilh\ (Hunter \et\ 1995), for their morphological type.

\subsection{Relationship of HI to \ha}

In Figure \ref{fighionha} we show contours of HI superposed on
our \ha\ image. We see that, with the exception of the two
faint \HII\ regions to the south of NGC 2363, star formation is
found within an observed HI column density exceeding 2$\times10^{21}$ \coldens.
The two extreme southern \HII\ regions sit within a
column density contour of half that.
In addition, the \HII\ regions are primarily associated with local
peaks in the HI. In particular, most of the \HII\ regions lie along
the eastern HI ridge.

In Figure \ref{figimvim} we compare the HI and \ha\ brightnesses
of individual pixels in the galaxy, where the \ha\ 
image has
been averaged 15$\times$15 pixels in order to approximately match the HI 
pixel size in the C$+$D-array map.
The x-axis of this figure goes from zero to an observed HI column density
of 5.2$\times10^{21}$ \coldens. The y-axis goes from an \ha\
luminosity of 7.2$\times10^{33}$ ergs s$^{-1}$ to 
1.4$\times10^{37}$ ergs s$^{-1}$, corrected for reddening
(1 count in the \ha\ image is 1.1$\times10^{33}$ ergs s$^{-1}$;
we have not converted this to a surface brightness by dividing by
the area of the pixel).
\ha\ is corrected for reddening assuming E(B$-$V)$_t$$=$E(B$-$V)$_f$+0.1$=$0.143.
With a Cardelli \et\ (1989) reddening curve, A$_{H\alpha}$$=$0.36.
There is a trend that the highest \ha\ luminosities are found
at the highest observed HI column densities. 

\subsection{Velocity Field}

The HI velocity field of NGC 2366 is shown in Figure \ref{figvel}.
We have fit the velocity field in the
following manner. We began by allowing all parameters to be variables
and fitting the entire field with a Brandt function. From the
resulting solution we fixed the center coordinates.
These are given in Table \ref{tabpa}.
We then fit the inner 150\arcsec\ radius of the velocity field
with solid body rotation and from that fixed the systemic velocity
at 99.1$\pm$0.2 \kms. 
We did this for inclination angles of
56\arcdeg\ and 72\arcdeg\ and the two resulting systemic velocities agreed
to 2 \kms.
(Swaters [1999] reports a systemic velocity
of 104 \kms, and Wevers \et\ [1986] report 98 \kms ). 
We then fit tilted ring models to annuli of 20\arcsec\ width out
to a radius of 460\arcsec,
allowing the inclination, position angle, and rotation speed
to vary and trying three different initial guesses. 
We found that the receding side, corresponding to the northeast portion
of the galaxy, produced more
stable solutions than the approaching or southwest side, and 
emphasis was placed on the receding side in determining the final solution.
In addition the fitting routine was unable to fit the innermost ring.
We also found that beyond a radius of 150\arcsec\ the 
position angle and inclination were approximately constant. 
The variations of position angle and inclination with radius are shown
in Figure \ref{figrot}. 
The position angle
and inclination were determined from the 150\arcsec$<$R$<$460\arcsec\ 
portion of the fit.
Finally, we fixed the position angle and inclination and fit
the rotation speed again for each annulus using annuli of 20\arcsec\
width to a radius of 460\arcsec\ and annuli of 40\arcsec\ width
beyond there to 620\arcsec.
We also used the higher resolution C-array velocity field to
examine the kinematics in the inner 3\arcmin\ with 10\arcsec\ rings.
The uncertainties were higher but the results were essentially the
same as for the C$+$D-array field.

The final position angle and inclination determined from the
HI kinematics are listed with the others determined from the optical
and HI morphology in Table \ref{tabpa}.
We determine an inclination angle of 65\arcdeg\
and a position angle of 46\arcdeg. 
This can be compared
to 58\arcdeg\ and 39\arcdeg, respectively, determined by Braun (1995)
from high resolution B-array VLA data; 65\arcdeg\ and 40\arcdeg\
determined by Wevers \et\ (1986); and 59\arcdeg\ and 42\arcdeg\
determined by Swaters (1999); thus, the agreement is good.
However, the
inclination and position angles determined from the HI velocity field
are not the same as those found from the optical or HI morphologies.
The morphologies of the stars and gas both give axis ratios that
imply an inclination of about 72\arcdeg, while the fit to the velocity
field suggests that the HI is 7\arcdeg\ less inclined. The position angle of 
the velocity field is also rotated further to the east than the
morphology of the stars or gas would suggest by 14--22\arcdeg.
These results indicate that either the HI distribution or the
kinematics deviate from axisymmetry. That is, the HI disk must
be either elongated or warped or the motions must be non-circular.

The resulting inclination-corrected rotation
curve is shown in Figure \ref{figrot}
(adopting the inclination of 65\arcdeg\ determined from the fit
to the velocity field),
and contours of the model rotation
curve are superposed on the velocity field in Figure \ref{figvel}.
The kinematical center of the gas is shifted 5\arcsec\ to the east
and 17\arcsec\ to the north of the center of the outer optical isophotes,
but these offsets are less than one beam-width.
One can see in Figure \ref{figvel} (and Figure \ref{figvdispha} below)
that the velocity field is distorted
to the southeast of the center of the galaxy, and this causes problems
in fitting the approaching side of the velocity field.
This region of the velocity field has no obvious connection with
any optical, integrated HI, or velocity dispersion feature. 
In fact, the anomalous region is outside the V-band contour shown
in Figure \ref{figvel}. Furthermore, the distortion in the velocity
field is not the S-shape of a warp (Bosma 1981).
There is also a 
distortion to the north of the galaxy center, and the isovelocity 
contours on the outer west side of the galaxy generally twist to become
oriented north-south at the edge of the HI distribution
(see Figures \ref{figvel} and \ref{figvdispha}).

In spite of all this the rotation curve shown in Figure \ref{figrot} is
fairly normal. It rises like a solid body for approximately
the inner 110\arcsec\ ($=$0.7R$_{25}$), rises less steeply to
$\sim$1.1R$_{25}$, and then levels off. At a radius
of about 340\arcsec\ ($=$2.2R$_{25}$), the rotation curve begins to fall
although the approaching and receding sides behave differently,
with the approaching side remaining approximately level.
The maximum rotation speed is 48 \kms.
Swaters (1999) gives a rotation speed of 60 \kms\ at his last
measured radius of 5.9\arcmin; the difference is consistent with
his lower value for the inclination.

\subsection{Bar Structure Revisited} \label{sect:revisit}

If the galaxy contains a stellar bar potential, we might expect
this to be reflected in velocity field anomalies in the vicinity of
the bar---in the isovelocity contours or changes in position angle
or inclination (for example, Pisano, Wilcots, \& Elmegreen 1998;
Koribalski \et\ 1996; Ondrechen \& van der Hulst 1989).
In the velocity field (Figures \ref{figvel} and \ref{figvdispha}) 
there is a hint of a position angle 
greater than 45\arcdeg\ for radii less than 1.5\arcmin. This could,
if real, be a bar effect. However, clear and obvious 
changes in the position angle or inclination
with radius that could correspond to the edge of a bar are missing,
although the resolution of the C$+$D-array data is too low to resolve the 
inner bar identified in Figure \ref{figallbars}. 
In addition, the
velocity field anomalies that we do see 
are not associated with the optical galaxy.
The velocity field
of the C-array data shows more variations everywhere, probably because
of the lower signal-to-noise.  And, in the C-array velocity field
there is an
anomaly at the southwest end of the inner bar, but this is also
where the supergiant \HII\ complex NGC 2363 is located and it is expected
to have an impact on the surrounding interstellar medium (ISM) there.
Thus, we see no velocity field anomalies that we can convincingly associate
with a bar structure.

A bar in NGC 2366 could be more subtle, however.  NGC 2366 is a later
type galaxy than is usually studied for bar flows, so we might not
see the same kinematic features that are found in other galaxies.
The LMC, for example, is a small system like NGC 2366, and it has a
clear optical bar, but this bar has virtually no signature in the HI
intensity and only a mild distortion of the inner HI velocity fields
(Kim et al. 1998). In this sense, it resembles NGC 2366.  In addition,
many late type galaxies tend to look elongated or irregular in outline
like NGC 2366, so it is possible that some of the observed ellipticity
in Figures \ref{figouterbar} and \ref{figallbars} is the result of an
intrinsically prolate structure.  If this is the case, then what should
we expect for non-circular perturbations in the velocity field?

Figure 4 in Athanassoula (1992) shows a model simulation that could be
like part of NGC 2366.  It is for a barred galaxy with no inner Lindblad
resonance (ILR), and it
shows a compressed central dust lane that runs almost like a straight
line from each end of the bar through the nucleus.  The bar compression
does not take the form of two offset dust-lanes with an inner spiral, as
in earlier-type galaxies with an ILR.  A straight dust lane in NGC 2366 could
account for the central ridge of star formation in Figure \ref{figallbars}
and for the thin inner HI and H$\alpha$ streaks in Figure \ref{fighionha}.
Another model with nearly straight dust lanes has a large bar axial ratio
(Fig. 6 in Athanassoula 1992).  Such axial ratios also result from the
lack of an ILR.

NGC 2366 has a solid body rotation curve in the inner part, as shown in
Figure \ref{figrot}. This implies that the value of $\Omega-\kappa/2$,
where $\Omega$ is the angular rotatino rate and $\kappa$ is the
epicyclic frequency,
is nearly zero there and makes it unlikely there is an ILR, which requires
$\Omega-\kappa/2$ to equal the rotation speed $\Omega$.  Thus, the Athanassoula
simulation seems relevant to NGC 2366.

The velocities in the Athanassoula (1992) solutions are parallel to the
bar in the inner regions, with a convergence and shock near the bar major
axis.  Such velocities would twist the HI minor kinematic axis away from
the optical minor axis in a manner like that seen in Figures \ref{figvel}
and \ref{figvdispha} (middle panel).

For example, the southwest side of NGC 2366 is approaching us, so the
counter-clockwise twist of the velocity contours in the inner region
(compared with the minor isophotal axis) indicates more of the same
approaching motions to the west of the center than in the east.  If NGC
2366 rotates counter-clockwise, so that the near side is in the northwest,
then this twist is consistent with a bar whose far end is toward the
south, and whose parallel approaching streaming motions are from the
far side to the near side in the western part of the disk. In this case,
the bar would be twisted slightly clockwise compared to the major axis of
the outer part of NGC 2366. Alternatively, if NGC 2366 rotated clockwise
(placing the near side in the southeast), then the observed twist of the
velocity contours would be consistent with a bar whose far end is toward
the north, and whose parallel streaming motions are from the far side to
the near side in the same western part of the disk. In this case also,
the bar would be twisted slightly clockwise from the major isophotal axis
of NGC 2366.  In both cases, we should look for an optical bar that is
slightly offset in the clockwise direction from the orientation of the
outer part of NGC 2366.
This offset might be seen in Figure \ref{figallbars}.
The contour level at 24.04 mag arcsec$^{-2}$ shows a twist of
the faint optical light in this sense, extending for the same radius
as the counter-clockwise twist of the HI velocity contours in 
Figure \ref{figvdispha}.

\subsection{Velocity Dispersion}

A false-color display of the velocity dispersion map is shown 
in Figure \ref{figvdispha}. 
Within the confines of the optical galaxy, the velocity dispersion of the
HI gas is mostly 12--20 \kms\ seen in the C$+$D-array map. This is high
compared to the canonical value of
10 \kms\ found in the quiescent gas of many galaxies including irregulars.
In the higher resolution C-array velocity dispersion map, however, we see that
most of the HI disk is in fact at 5--10 \kms\ dispersion but that
the HI ring, including regions without star formation, has dispersions
of 10--17 \kms.

Figure \ref{figvdispha}
shows a comparison of the velocity dispersion to the V-band and
\ha, the velocity field, and the integrated HI by superposing
contours in three displays.
One can see that there are regions of high velocity dispersion,
20--30 \kms, scattered around the HI disk.
One would expect the
velocity dispersion of the neutral gas to be high in regions of star
formation. In the supergiant \HII\ regions in particular, 
where there is intense star
formation, the ionized gas itself is known to have high velocity dispersions
(Hunter 1982; Hunter \& Gallagher 1990, 1992, 1997).
However, Figure \ref{figvdispha} shows that the regions with
the highest velocity dispersions in HI are found outside the regions of star
formation and beyond the bright part of the optical
galaxy altogether. In fact the high velocity dispersions
are not associated with any obvious feature in the optical or HI.
The only exception is that an ``arm'' of higher velocity dispersion 
curves around the northern and western sides of the ``knob'' of HI
in the northeast part of the galaxy (see bottom panel of Figure 
\ref{figvdispha} and Figure \ref{figgrayfull}).

Some regions of high velocity dispersion occur where the velocity
contours crowd together, as illustrated in the middle panel of Figure
\ref{figvdispha}. This is true particularly in the northwest region, where the
contours are vertical (7$^h$ 23$^m$ 13.4$^s$, 69\arcdeg\ 21\arcmin\ 07\arcsec).
In this case, part of the high dispersion comes from
unresolved systematic motions.
However, in other cases the contribution from rotational motions is small.
For example,
profiles of the region of high velocity dispersion to the east
of the center of the galaxy (7$^h$ 24$^m$ 00.6$^s$, 69\arcdeg\ 18\arcmin\
32\arcsec)
are shown in Figure \ref{figplcubeast}.
The line profiles in these regions are simply very broad and in
some places double peaked.

The regions of high velocity dispersion not significantly affected
by the velocity gradient, especially the 
spiral-like features in the outer parts of the disk, are more difficult
to explain. They could have a turbulent origin.  Figure \ref{vdcolumn}
shows scans of column density as dashed lines and velocity dispersion
as solid lines, taken across the entire disk from east to west.
Each scan is averaged
in vertical bands that are one beam FWHM wide. 
They go through three main regions of high velocity dispersion,
which are at the indicated positions:
E ($7^h 24^m 00.6^s$, $69^\circ 18^\prime 32^{\prime\prime}$),
N ($7^h 23^m 38.9^s$, $69^\circ 20^\prime 52^{\prime\prime}$;
includes the northwest region as well),
and
SW ($7^h 22^m 58.4^s$, $69^\circ 16^\prime 57^{\prime\prime}$).
The figure
shows a tendency for the velocity dispersion to increase where the column
density decreases. There is nearly a one-to-one correspondence between
these opposing variations in the central regions of the galaxy. This
suggests a process where there is some pressure or wave-like communication
between the regions. Low density regions have greater wave or turbulent
motions than high density regions, in a manner analogous to the turbulent
motions in individual clouds.

Another factor that can affect line profiles is low signal-to-noise, and
the regions with high velocity dispersion are also regions of low
HI column density. It is claimed that noisier profiles, such as one might
expect in low intensity regions, will appear broader. If this were
the explanation of the high $c$ regions that we see here, then we would
expect all regions in the outer parts of the galaxy to have high $c$. 
Yet, in Figure \ref{figvdispha} we see a few selected regions only with
anomolously high $c$. Furthermore, Figure \ref{figplcubeast} shows
the profiles in one of these regions, and they are clearly broad even
though noisy. Thus, while this effect may be entering in at some level,
we believe that these regions of anomolously
high $c$ and the trend we see in Figure \ref{vdcolumn} are real.

This anti-correlation between velocity dispersion and HI column density
that we find in NGC 2366
is different from what is seen in the Scd galaxy IC 342.
There, over most of the galactic disk $c$ is highest
where the gas column density is highest (Crosthwaite, Turner, \& Ho 2000).
The exception in IC 342 is the center where $c$ peaks and the HI column
density drops. Crosthwaite \et\ suggest that in the center of IC 342
some additional
gas component is contributing to the total gas surface density.
Elsewhere in IC 342, the correlation between the surface density and velocity
dispersion suggests there is a local source of turbulent energy 
from gas compression in a density wave or local spiral instability. 
In NGC 2366, there are apparently no analogous spiral instabilities.

However, there is another situation in which there is an
anti-correlation between a velocity dispersion and HI column density.
That is in the irregular galaxy NGC 4214. In a comparison of
\ha\ and HI velocity fields in that galaxy, Wilcots \& Thurow (2001) 
found that the highest \ha\ velocities are those of the diffuse
ionized gas, often in regions of low HI column
density and holes. They argue that these are the signatures of
outflows fueled by surrounding concentrations of massive stars
and accelerated by the steep density gradients into holes.
However, this seems unlikely to explain the regions of high $c$ in
the HI in NGC 2366. Although the regions of high $c$ in NGC 2366 are located
in the outer parts of the galaxy where there is a general density
gradient, there are not the
concentrations of large numbers of massive stars nearby that are seen in 
NGC 4214 to supply the mechanical energy input. The exception is
NGC 2363, of course, but there are pockets of high $c$ in NGC 2366
located on the other side of the galaxy from NGC 2363. Furthermore,
in this scenario the high velocity dispersion is in the {\it ionized}
gas, not the neutral gas, since the massive stars are ionizing
the medium as well.

\subsection{HI and NGC 2363}  

The current star formation activity of NGC 2366 is dominated by
the supergiant \HII\ complex NGC 2363, the scale of which is
truly remarkable
(Kennicutt, Balick, \& Heckman 1980). Star formation appears
to have been continuing over a period of time up to 10 Myr
throughout the complex.
There is spectroscopic evidence of the presence of Wolf-Rayet
stars, indicating an age of 3--5 Myr, if they
are normal evolved massive stars,
near the brightest part of the nebula
(Drissen, Roy, \& Moffat 1993; Gonz\'alez-Delgado \et\ 1994).
Star formation over periods of order 10 Myr is also what is observed
in large \HII\ complexes in the LMC such as 30 Doradus or Constellation III
(see, for example, discussion by Dolphin \& Hunter [1998] 
and references therein).

NGC 2363 also has a system of \ha\ arcs that form a partial
shell 835 pc long along the northeast perimeter of the \HII\ complex 
(Figure \ref{fign2363}).
If they represent a shell centered on the brightest nebula
in the complex, the diameter of the shell would be 1500 pc
(Hunter \& Gallagher 1990).
Spectroscopy shows that the arcs are photoionized 
(Hunter \& Gallagher 1997).
In addition to the arcs, there are diffuse filaments of ionized
gas 2--5 times lower in surface brightness that extend from the 
western-most end of the arcs to the northwest away from the
nebula (Hunter, Hawley, \& Gallagher 1993). 
These filaments are of order 770 pc long and 60 pc wide.
They extend to 1130 pc from the center of NGC 2363.

Furthermore, high velocity ionized gas with FWHM of 38 \kms\
and FWZI of 190 \kms\ is associated with the nebula;
the ionized gas alone has a kinetic energy
of 9$\times10^{50}$ ergs s$^{-1}$
(Hunter 1982, Roy \et\ 1992, Hunter \& Gallagher 1997, Martin 1998).
Roy \et\ (1991) argue that there is a 285 pc diameter
bubble of gas expanding at 45 \kms\ in the nebula.
All of this motion of the ionized gas is most certainly
a consequence of the concentration of massive stars in the complex.
Evidence is seen for at least one supernova remnant
(Yang, Skillman, \& Sramek 1994).

We have examined the HI within and around NGC 2363
to look for any relationship to the ionized gas structures
and unusual activity in the gas.
In Figure \ref{fign2363} we show our \ha\ image of NGC 2363.
One can see the arcs of the partial shell curving around the \HII\ region
to the northeast and the faint, diffuse filaments
trailing away from the nebula to the northwest. There is
also diffuse \ha\ emission to the west between NGC 2363 and the 
giant \HII\ region to the west (numbered III by Drissen \et\ 2000).
Contours are superposed on the \ha\ image from the higher resolution
C-array integrated HI map. One can see that the \HII\ region is
offset from the HI peak, as is commonly seen
(for example, Sextans A: Hodge, Kennicutt, \& Strobel 1994). The \ha\ peak
is 18\arcsec\ (300 pc), about one HI beam-width, 
to the west of the HI peak. The faint, diffuse filaments to the
northwest are entering a region that is a local HI minimum
as would be expected if they are part of a champagne flow
away from the gas cloud.
The ionized gas arcs to the northeast, on the other hand, cross the HI ridge
and there are two minor HI peaks located along the
shell as well. Where the ridge of HI ends as one moves to the
northwest along the arcs is where the arcs
end and the faint, diffuse filaments begin.

Kinematically there is some anomaly in the C-array
velocity field at NGC 2363. 
This can be seen in Figure \ref{fig2363vel} where we show contours
of the velocity field superposed on our \ha\ image. In the
vicinity of the \ha\ arcs on the northeast side of NGC 2363
the velocities of the HI are lower than one would expect 
except for a small pocket were the velocities
are higher. However, the magnitude of these variations is of order
5 \kms, which is less than the velocity dispersion of the gas.
In Figure \ref{fign2363plcub} we show line profiles across
the NGC 2363 region and immediate vicinity. 
Although a few line profiles are double, most are not.
Furthermore, the FWHM of the line profiles in the \HII\ region,
in the arcs, and in the diffuse filaments are comparable
to what is seen throughout the galaxy.

\subsection{Shells}

Holes and shells have been found in the HI distributions of
various irregular galaxies.
Most shells within the optical galaxy
are believed to be the consequence of the energy input from 
massive stars that have recently formed, although
non-stellar explanations have been proposed for larger holes,
especially those outside the optical galaxy (Wada, Spaans, \& Kim 2000).
Shells are seen in the HI distributions of galaxies such as
DDO 50 (Puche \et\ 1992), IC 10 (Wilcots \& Miller 1998), and
the LMC (Kim \et\ 1999).
In those galaxies the shells are apparent in the integrated
HI maps. We do not see evidence of shells in our integrated map
of NGC 2366, nor in a careful examination of
channel maps and position-velocity
plots.
Of the 51 shells
catalogued in DDO 50 by Puche \et\ 
the diameter ranges from 140 to 1900 pc with an
average of 730$\pm$440 pc. The average size in IC 10
is 160$\pm$40 pc (Wilcots \& Miller 1998). 
If shells were present in NGC 2366, our resolution
of 300 pc$\times$220 pc (C-array map) should have allowed us to 
detect the largest structures found in DDO 50. However, NGC 2366
could contain numerous more modest-sized shells that we would
not be able to resolve with these data.

\subsection{The Suggestion of an Interaction Considered}

Drissen \et\ (2000) have suggested that \HII\ region number III to the west
of NGC 2363 and the ridge of gas that it sits on are actually a satellite
galaxy now interacting with NGC 2366. They base this idea on the
age sequence running from east to west through the NGC 2363 
star-forming complex and suggest that the interaction initiated 
star formation in this region which then spread in a sequential manner.
However, as we have seen, the HI, although having significant peculiarities,
is not as peculiar as one
might expect in the case of an on-going merger. 
Morphologically, the two ridges of gas
deproject into a lumpy ring of gas that is not particularly unusual
in appearance. Further, in this ring the gas complex associated with region III
does not stand out as particularly unusual compared
to the NGC 2363 complex.
Kinematically, we have noted that there are large-scale peculiarities:
the position angle of the kinematical axis is rotated from those
of the optical and HI morphology, there is gas extending southwest
of NGC 2363 at unusual velocities, and the isovelocity contours
on the west side of the galaxy rotate to become almost north-south.
However, we do not see evidence
of gas at unusual velocities specifically associated
with region III.
Futhermore, similar peculiarities are found in many other galaxies.
Thus, while it is possible that the velocity field peculiarities do indicate
that NGC 2366 has been disturbed by an
outside force sometime in the past, we do not feel compelled by
the data to assign \HII\ region III outsider status at this time.

\section{Star Formation in NGC 2366}

There is evidence for an
empirical dependence for local star formation activity within a galaxy
that involves a threshold.
In a thin, differentially-rotating disk composed of pure gas (Safronov
1960) or stars only (Toomre 1964), there is a critical surface density
\sigcrit\ above which the disk is unstable to ring-like perturbations in
the radial direction.
Kennicutt (1989) calculated the critical gas density in this
model (Quirk 1972) for a sample of normal Sc spiral
galaxies.  He found that the ratio of observed gas density \siggas\ to
critical gas density \sigcrit\ typically exceeds 1 at mid-radius in the
optical disk, and then falls with increasing distance from the center of
the galaxy; star formation was not detected beyond the point where
\sigrat\ dropped below 0.7$\pm$0.2. He concluded that in Sc galaxies 
places where \sigrat$<$0.7 are too stable to form stars; interior to this,
the gas density exceeds the local threshold for star formation.

Motivated by the apparent success of the gravitational instabilities
model for spiral galaxies, we
explored the applicability
of several models to a small sample of irregular galaxies
(Hunter, Elmegreen, \& Baker 1998;
see also Hunter \& Plummer 1996, Meurer \et\ 1996, van Zee \et\ 1997).
These models included
(1) thin-disk, single-fluid gravitational instabilities,
(2) thick-disk, two-fluid (stars+gas) gravitational instabilities, (3)
three-dimensional gravitational instabilities based on a comparison
between the local gas volume density and the local galactic virial
density, (4) gravitational
instabilities limited by the shear time, which uses a modified
\sigcrit\ based on the Oort shear constant $A$ instead of the
epicyclic frequency $\kappa$,
(5) the two-phase thermal properties of the ISM, and (6) any of
a variety of models, presumably based 
on triggering or feedback, in which the star formation rate depends on
the local stellar density.  In addition, we also considered all of
these models in a second case where the relative density of dark
matter is high in the disks of irregular galaxies, as motivated by the
high relative dark matter densities overall (e.g.  Carignan \&
Beaulieu 1989).
The resulting critical densities for cloud formation for each of
these models were compared
with the observed gas surface densities.
The ratio of observed to critical gas density as a function of radius
was also compared with
the azimuthally-averaged current star formation rate
and with the V-band stellar surface brightness,
a tracer of star formation integrated over a timescale of $\sim$ 1 Gyr.

We found that the thin-disk, single-fluid model that works so well for
spirals does not work for irregulars.  
We found that the ratios of observed to critical gas densities are
lower in irregulars than in spiral galaxies by a factor of $\sim2$
in all cases but the shear-regulated model, suggesting that the gas
is highly stable on large scales against cloud formation.
The
absolute criterion based on this model, therefore, predicts that star formation
should not be occurring at all in irregulars.  In addition, 
the relative criterion
as a function of radius does not predict the observed star formation
edge in the galaxies, nor does it predict
the variation in star formation rate from galaxy to galaxy.  

In Figure \ref{figsig} we show this failure of the models for
NGC 2366 as well. There we plot the deprojected surface densities
\siggas\ and the \sigcrit\ 
for the thin rotating disk gravitational instability model.
The values of \siggas\ have been determined from \sighi, assuming
the kinematically-determined inclination
of 65\arcdeg. 
If one were to use the inclination of 72\arcdeg\ suggested by
the morphology, the dominant effect is the $\cos(i)$ in
deprojecting the column densities. In that case the
measured surface densities decrease by a factor of $\sim$0.7.
In addition \sighi\ has been 
corrected for He (0.34$\times$\sighi).
Including molecular gas is far more difficult; few irregular galaxies
have been adequately mapped, so the spatial distribution of H$_2$ is not
generally known.  Instead, we took a globally averaged
M$_{H_2}$/M$_{HI}$ ratio and simply multiplied the HI everywhere by this
in order to statistically include H$_2$.  For the M$_{H_2}$/M$_{HI}$
ratio we used an average between the SMC, where molecular gas is 7\% of
the HI gas content (Rubio \et\ 1991), and the LMC, where the ratio is
30\% (Cohen \et\ 1988).  That is,
the H$_2$ content was taken to be 0.18$\times$\sighi. 
This correction for the presence of H$_2$ is
clearly unsatisfactory, but at present there is no better information on
these galaxies.  Thus, we took \siggas\ to be 1.52$\times$$\Sigma_{HI}$.

\sigcrit\ was determined from
${\Sigma_c} = {{\kappa c}\over{3.36G}}$, where $\kappa$ is the epicyclic
frequency and $c$ is the velocity
dispersion of the gas.  We determined $c$ as a function of radius
by integrating the velocity dispersion map in ellipses of 25\arcsec\
width and using a position angle of 46\arcdeg\ and an inclination angle
of 65\arcdeg. The mean velocity dispersion in these ellipses, 
determined from the C$+$D-array 
map, is about 16 \kms\ out to a radius of 5\arcmin, and declines
slowly from there to 10 \kms\ at a radius of 11.7\arcmin.
The values of \sigrat\ in NGC 2366
are typical of the irregulars in our previous study: The maximum in
\sigrat, 0.5, is still below the star formation cut-off value of
0.7 found by Kennicutt (1989) for spirals.

These failures of the models suggest that other processes are
important for cloud and star formation in irregulars, 
leading us to suggest more local effects.
An obvious mechanism is sequentially triggered star formation
driven by the mechanical energy
input from concentrations of massive stars resulting in gas density
irregularities.
Another process
is random gas compression from turbulence.
Kinematics in Im systems are often dominated
by random gas motions which can contribute significantly to the pressure
(for example, Gottesman \& Weliachew 1977;
Tully \et\ 1978; Huchtmeier \et\ 1981; Sargent, Sancisi, \& Lo 1983;
Lo, Sargent, \& Young 1993; Feitzinger \et\ 1981), and in some cases rotation
can be insignificant compared to random motions (Young \& Lo 1996).
Both of these
processes have a sensitivity to the critical density like the
gravitational instability model, but do not involve large-scale
instabilities directly.  They could in fact dominate the star
formation process when spiral arms are not present or are weak
(see also Thornley \& Wilson 1995).
Further evidence that local inhomogeneities in the gas may dominate
the star formation process come from 
the observations that \HII\ regions
are found near peaks in the gas even when the azimuthally-averaged
gas is below the
critical threshold density (van der Hulst \et\ 1993;
van Zee \et\ 1997;
Meurer \et\ 1998).

Because of the apparent dominance of local inhomogeneities in the gas 
distribution, we have
determined the maximum \siggas\ at the positions of the \HII\ regions
in NGC 2366, \sighii. 
As for \siggas, \sighii\ uses an inclination angle of 65\arcdeg\ to
convert from column density to surface density and is corrected 
to account for the presence of He and, statistically, H$_2$.
We have determined the maxima using the
C-array maps for the higher spatial resolution (300 pc$\times$220 pc)
they give
and have looked for the HI peaks within a beam area or the diameter of
the \HII\ region, whichever was larger. For most \HII\ regions
the beam-size is
larger than the \HII\ region diameter. This does not take into account
the possibility that the \HII\ region is offset from the gas cloud
out of which it has formed, nor can it account for destruction of the
cloud since the formation of the \HII\ region.
The values of \sighii\ 
at each \HII\ region are shown in Figure \ref{figsig} along with
the \sigcrit\ from the thin-disk instability model and the azimuthally-averaged
\siggas.
We see that the \sighii\ cover a relatively small range, 
18--36 M\solar\ pc$^{-2}$, except in the center ($<$0.5R$_{25}$)
and outer ($>$1.75R$_{25}$) galaxy where \sighii\ are lower.
In fact, the \sighii\ are comparable to the annular \siggas\
in the inner 0.5R$_{25}$ of the galaxy, but 
beyond $\sim$0.75R$_{25}$ the \sighii\ are roughly 
comparable to \sigcrit.
The transition region corresponds roughly to the radius at which
the rotation of the galaxy ceases to be that of a solid body
(inner vertical dashed line).
Also, star formation does not extend very much beyond the radius
where the rotation curve begins to fall 
(outer vertical dashed line).

The rise and fall of \sighii\ is similar to the rise and fall of
the limiting tidal density in the disk midplane. The tidal density is
\begin{equation} \rho_T=-{{3\Omega R}\over{2\pi G}} {{d \Omega}\over{dR}}
\end{equation} 
for angular rotation rate $\Omega$ and radius $R$.
In the rising part of the rotation curve, $d\Omega/dR$ is small so
$\rho_T$ is small, and in the flat part of the rotation curve, $\rho_T$
decreases as $1/R^2$.  To compare this with the observations, we have
to convert $\rho_T$ into a tidal-limit 
surface density, $\Sigma_T$.  We
multiply $\rho_T$ by twice the scale height, $H= c/(2\pi G\rho_T)^{0.5}$
to get
\begin{equation} \Sigma_T={{\Omega c}\over{\pi G}}
\left(-3d\ln\Omega/d\ln R\right)^{0.5} \end{equation} 
for velocity dispersion
$c$. This quantity, corrected for the presence of He and H$_2$
as for \siggas,
is plotted in Figure \ref{figsig} as crosses.
The general agreement of $\Sigma_T$ with \sighii\ indicates that the peak
column densities near the HII regions are not particularly dense, but
only at about the tidal limit.  Lower density clouds should not bind
themselves together well.  The tidal limit in the flat part of the rotation
curve is higher than the peak column density, but this is probably a
result of our not including stars in the expression for $H$.  If we write
$H= c/\left[2\pi G\left(\rho_T+\rho_{stars}\right)\right]^{0.5}$, 
then the tidal limit
equals the observed \sighii\ if $\rho_{stars}\sim0.6\rho_{gas}$.
Stars in the inner disk lower $\Sigma_T$ too, making the clouds there
relatively more self-bound. 

Various people have suggested that, rather than a varying critical
density stability criterion,
there is a simple constant and universal density threshold for star formation
(Gallagher \& Hunter 1984; 
Skillman 1987; 
Taylor \et\ 1994; 
Meurer \et\ 1996; 
Hunter \et\ 1998; 
van Zee \et\ 1998). 
Most estimates for this threshold are observed column densities of
$\sim10^{21}$ \coldens, which 
corresponds to an HI surface density (corrected for inclination of
the galaxy) of 3 \sigdens\ in NGC 2366 or \siggas\ of
5 \sigdens\ corrected for He and H$_2$. We had also estimated a
critical gas density of 3 \sigdens\
from azimuthally-averaged \siggas\ for a sample of irregulars.
In NGC 2366 we see that all of the star formation is taking place
at surface densities
\sighii$>$6 \sigdens\ or at an azimuthally-averaged \siggas$>$3 \sigdens,
and the bulk of it is taking place at $>$15 \sigdens. The data for NGC 2366
are not inconsistent with this simple, universal threshold.

Additional support for this universal threshold comes from comparing
peak column densities in galaxies with different levels of
star formation activity. The caveat, as always, is that the measured
column densities depend on the beam-size of the observations relative to
the size of the galaxy. Nevertheless, they can give us some indication
of differences between galaxies.
In our larger sample of irregulars assembled for purposes of studying
star formation processes, there are four galaxies with no measurable
star formation activity (determined from \ha\ images) and with published
HI maps given with contour levels. In these four galaxies the highest
observed HI contours are column densities of 1--7$\times10^{20}$ \coldens\
(LGS3--Young \& Lo 1997; M81dwA--Sargent \et\ 1983;
F564-V3 and F567-2--de Blok, McGaugh, \& van der Hulst 1996).
These column densities correspond to deprojected HI surface densities
of 0.8--5 M\solar\ pc$^{-2}$ for these galaxies.
There are also three galaxies in this sample with star formation rates
per unit area less than $10^{-4}$  \sfrunit, the lowest non-zero rates measured
in this sample, and they have HI contour levels shown to 7--24$\times 10^{20}$
\coldens\ (DDO 52--Simpson 1995; DDO 69--Young \& Lo 1996, note the factor
of two difference in N$_{HI}$ for the two beam sizes that they give;
DDO 216--Lo \et\ 1993). 
These column densities correspond to HI surface densities of 2--10
M\solar\ pc$^{-2}$.
Of three galaxies with somewhat
higher star formation rates---$10^{-3}$--$10^{-4}$ \sfrunit, the highest
HI contours are 7--14$\times10^{20}$ \coldens\ (DDO 105--Broeils 1992;
F563-1 and F565-V2--de Blok \et\ 1996).
These column densities correspond to HI surface densities of 3--5
M\solar\ pc$^{-2}$.
The picture is not clean but suggests that something like
an HI surface density of $\sim$2--5 M\solar\ pc$^{-2}$, or
a \siggas, corrected for He and H$_2$, of 3--8 M\solar\ pc$^{-2}$
could be a universal cutoff for massive star formation.
Of course, these estimates are based on looking at the contours that
enclose the bulk of the star formation and do not refer to the peaks
in \siggas.

There is good reason for an absolute minimum surface density threshold
in irregular galaxies, in addition to whatever other
thresholds there might be, and this is the requirement of a cool phase
of HI as a precursor to star formation. The existence of a
cool phase depends on the pressure and radiation field. For
normal radiation fields in star-forming disks, the pressure
has to exceed several times 10$^3$ K cm$^{-3}$ for the HI
to allow an equilibrium cool phase. This pressure
comes entirely from the disk self-gravity, and depends on the
square of the column density. 
For a \siggas\ of 8 M\solar\ pc$^{-2}$,
the pressure is $\left(\pi / 2\right) G \Sigma_g^2\sim
2000$ K cm$^{-3}$.  Lower column-density regions should have their
HI only in the warm, low-density phase (Elmegreen \& Parravano 1994;
Braun 1997).

While it is clear that star formation is dependent on local gas peaks
to form stars, to some extent this is a circular argument. We know that star
formation begins with the formation of a gas cloud, and so it is not
at all surprising that local peaks in the gas are where we find \HII\ regions.
Once a gas cloud forms and reaches a certain density, it forms stars.
The larger question is what causes those clouds to form in the first place
in irregular galaxies. Large-scale gravitational instabilities
do not seem to operate with the same vigor as in 
giant spiral galaxies, because the gas
has a lower ratio of $\Sigma_g/\Sigma_{c}$ by a factor of 2.
Another mechanism could take over in the absence of
strong self-gravity, and that is the rearrangement of gas
by the input of mechanical energy from concentrations of massive stars
in previous episodes of star formation. This is the 
Stochastic Self-Propagating Star Formation (SSPSF) model of
Gerola, Seiden, \& Schulman (1980;
Wada, Spaans, \& Kim 2000),
based on complex feedback and triggering scenarios.

In NGC 2366 Aparicio \et\ (1995) argue that there is no large-scale
propagation between regions seen in color-magnitude 
diagrams of the stellar populations. However, their CCD images 
did not include the
redder regions of the galaxy that we see in Figure \ref{figdivbv}.
The redder stellar population situated in the center of the HI ring
which now hosts most of the star formation activity of the galaxy
does suggest some sort of propagation from a central point outward.
This is similar to what is seen in the irregular galaxy Sextans A 
($=$DDO 75): The center of Sextans A, where there is an HI minimum,
is dominated by older stars
(Hoessel, Schommer, \& Danielson 1983;
Aparicio \et\ 1987; Dohm-Palmer \et\ 1997).
Outside of the center, the bulk of the star formation is taking
place in regions associated with two large HI complexes
(Skillman \et\ 1988).

Without a previous generation of stars, star formation still had
to begin somehow, and perhaps this is where gravitational instabilities
contribute. Even with a low ratio $\Sigma_g/\Sigma_{c}$, instabilities
can still operate when there is energy dissipation (Elmegreen 1991;
Elmegreen 1995; Gammie 1996), but they operate relatively slowly, at
the viscous rate.  This type of slow instability might be more important
for irregulars than spirals, because the lack of competition from shear
in irregulars means there is a lot of time available. Indeed, most of
the visible disk of NGC 2366 has virtually no shear, so cloud-forming
instabilities can take a very long time to develop.

Dohm-Palmer \et\ (1998) in a study of the stellar populations in the
irregular galaxy DDO 155 ($=$GR8) noted that the recent star formation
is concentrated near the main HI complexes. They argue that the HI
complexes are long-lived objects and that star formation takes place
within them over hundreds of Myr without destroying the complexes.
The giant complex NGC 2363
could be long-lived too, but the similarity of the peak HI density 
to the critical tidal density in Figure \ref{figsig}
implies that NGC 2366, like the
other star-forming regions, is not tightly self-bound on a large
scale.  
A good test of this hypothesis would come from the
star formation history of
this complex.  If there is a population of stars dating back
several hundred million years that is exclusively in NGC 2363,
then the complex may be relatively long lived.

\section{Summary}

We have presented deep UBVJHK\ha\ images and HI maps of NGC 2366.
We see that the galaxy is a boxy-shaped structure at relatively
high inclination.  The optical morphology suggests several bar
structures including one in which 
the whole galaxy is
a bar.
However, there is only a weak observational signature of a bar potential in 
NGC 2366.
There is an asymmetrical extension of stars
along the southwest end of the major axis of the galaxy, 
and this is where the furthest star-forming
regions are found, at a radius of 1.3R$_H$. The colors of the galaxy
are approximately constant with radius when averaged azimuthally,
but show large-scale structure in ratio images.
The center of the HI ring and a low surface brightness region to the
northeast appear to be redder than the rest of the galaxy.
The optical surface photometry
is fit with an exponential disk that also applies to the near-infrared
surface photometry. The scale length and central surface brightness
of the disk are normal for late-type galaxies.

The star formation activity of the galaxy is dominated by the supergiant
\HII\ complex NGC 2363 that contains 70\% of the total \ha\ luminosity
of the galaxy. The global star formation rate for NGC 2366 is, however,
only slightly elevated 
for an Im system, and there is enough gas to last another one
or two Hubble times at the current rate. 
The star formation activity drops off with radius
approximately as does the starlight until the outer part of the galaxy 
is reached and then it drops faster.

We detected HI to a radius of 2.3R$_H$ at 1.7$\times10^{19}$ \coldens.
The integrated HI shows two ridges running parallel to the major axis
that when deprojected appear as a ring of diameter 5.5 kpc with
a central depression and a somewhat redder stellar population, 
as is seen in some other irregulars. 
Most of the star formation is found along the eastern of the two ridges,
but one major \HII\ complex is on the western ridge.
The HI is not centered symmetrically on the optical galaxy.
The azimuthally-averaged surface density of HI is comparable to that of other
irregulars in the inner part of the galaxy but drops off slower and
extends further in the outer parts.
Globally, NGC 2366 is more gas-rich than other irregulars.

The kinematics of the HI are not those of a well-behaved
rotating disk. The velocity field to the southeast of the center of
the galaxy and along the western edge
especially deviate from a model rotating disk.
There could be a weak kinematic signature
of a bar tilted clockwise relative to the major axis of the galaxy by
a few tens of degrees. 
The inclination and position angles derived from the kinematics differ
from those derived from the optical and HI morphology by 7--22\arcdeg.
There are regions of unusually high velocity dispersion, but these
regions are not associated with the optical galaxy or features
in the HI.
Most regions of high velocity dispersion correspond instead to regions of slightly
depressed HI column density, as if there is some continuity from region to
region in the turbulent pressure and wave motions of the interstellar medium. 
Within the
confines of the optical galaxy, the velocity dispersion is higher in
the HI ring than in the surrounding disk.
HI clouds have properties like those of clouds in other luminous irregulars.
We did not detect any shell structures in the HI.
The HI around the star-forming complex NGC 2363 is fairly unremarkable
with a small velocity anomaly in the vicinity of the \ha\ arcs,
and the large ionized gas shell appears to be perpendicular to
the HI ridge containing the complex.

Like in other disk galaxies, the gas in NGC 2366 is lumpy and star-forming regions
are associated with these lumps. \HII\ regions are found where the gas 
surface densities locally exceed 6 \sigdens. What causes these clouds to form
in the first place?
NGC 2366, like other irregulars, has low gas densities relative to the
critical gas densities of gravitational instability models, so large-scale
gravitational instabilities would operate slowly or not at all.  Such a
slowness of cloud-forming instabilities may not be as much of a problem
for irregular galaxies as it is for giant spirals because of the lack of
shear in the irregulars. When shear is low, there is little competition
to the growth of giant cloud complexes following turbulent energy
dissipation.

\acknowledgments

DAH is very appreciative of the travel support and hospitality
of the Kapteyn Astronomical Institute that enabled her to
travel to the Kapteyn Laboratory in Groningen where she received
invaluable tutoring in the analysis of HI-interferometric data
and the use of GIPSY. We thank A.\ Tomita for a critical reading
of the manuscript.
We appreciate the use of Ohio State University's Fabry-Perot and
the Ohio State 
Infrared Imager-Spectrograph 
and Mark Wagner for making them work for us.
Support to DAH for this research came from the Lowell Research
Fund and grant AST-9802193 from the National Science
Foundation; support to BGE came from grant AST-9870112 from the
National Science Foundation.

\clearpage

\begin{deluxetable}{rrrrr}
\tablecaption{Continuum Objects in the Field. \label{tabcont}}
\tablewidth{0pt}
\tablehead{
\colhead{} & \colhead{Flux}   & \colhead{RA} & \colhead{DEC} 
& \colhead{FWHM\tablenotemark{a}}\nl
\colhead{Object}  & \colhead{(Jy)} & \colhead{(1950)} & \colhead{(1950)} 
& \colhead{(\protect\arcsec)}
}
\startdata
NGC 2363 & 0.024 & 7$^h$ 23$^m$ 25.2$^s$ & 69\arcdeg\ 17\arcmin\ 30\arcsec & \nodata \nl
1 & 0.053 & 7$^h$ 19$^m$ 30.8$^s$ & 69\arcdeg\ 20\arcmin\ 15\arcsec & 38 \nl
2 & 0.009 & 7$^h$ 20$^m$ 43.0$^s$ & 69\arcdeg\ 23\arcmin\ 25\arcsec & 43 \nl
3 & 0.012 & 7$^h$ 20$^m$ 59.1$^s$ & 69\arcdeg\ 15\arcmin\ 18\arcsec & 35 \nl
4 & 0.027 & 7$^h$ 21$^m$ 05.5$^s$ & 68\arcdeg\ 59\arcmin\ 02\arcsec & \nodata \nl
5 & 0.022 & 7$^h$ 21$^m$ 54.0$^s$ & 69\arcdeg\ 25\arcmin\ 28\arcsec & 41 \nl
6 & 0.012 & 7$^h$ 22$^m$ 04.4$^s$ & 69\arcdeg\ 30\arcmin\ 01\arcsec & 40 \nl
7 & 0.002 & 7$^h$ 22$^m$ 29.2$^s$ & 69\arcdeg\ 20\arcmin\ 57\arcsec & \nodata \nl
8 & 0.179 & 7$^h$ 22$^m$ 39.2$^s$ & 68\arcdeg\ 58\arcmin\ 57\arcsec & 38 \nl
9 & 0.012 & 7$^h$ 22$^m$ 40.6$^s$ & 69\arcdeg\ 36\arcmin\ 49\arcsec & 39 \nl
10 & 0.069 & 7$^h$ 23$^m$ 14.9$^s$ & 68\arcdeg\ 57\arcmin\ 54\arcsec & 37 \nl
11 & 0.052 & 7$^h$ 23$^m$ 14.9$^s$ & 68\arcdeg\ 56\arcmin\ 50\arcsec & 43 \nl
12 & 0.008 & 7$^h$ 24$^m$ 35.2$^s$ & 68\arcdeg\ 59\arcmin\ 13\arcsec & 39 \nl
13 & 0.047 & 7$^h$ 24$^m$ 55.3$^s$ & 69\arcdeg\ 35\arcmin\ 21\arcsec & 36 \nl
14 & 0.014 & 7$^h$ 25$^m$ 07.3$^s$ & 69\arcdeg\ 12\arcmin\ 00\arcsec & 36 \nl
15 & 0.012 & 7$^h$ 25$^m$ 25.7$^s$ & 69\arcdeg\ 16\arcmin\ 24\arcsec & 38 \nl
16 & 0.005 & 7$^h$ 25$^m$ 28.3$^s$ & 69\arcdeg\ 11\arcmin\ 20\arcsec & 35 \nl
17 & 0.044 & 7$^h$ 26$^m$ 22.7$^s$ & 69\arcdeg\ 25\arcmin\ 49\arcsec & 37 \nl
18 & 0.010 & 7$^h$ 26$^m$ 20.3$^s$ & 69\arcdeg\ 31\arcmin\ 01\arcsec & 31 \nl
19 & 0.085 & 7$^h$ 27$^m$ 14.0$^s$ & 69\arcdeg\ 06\arcmin\ 01\arcsec & 37 \nl
20 & 0.077 & 7$^h$ 28$^m$ 00.8$^s$ & 69\arcdeg\ 15\arcmin\ 33\arcsec & 49 \nl
\enddata
\tablenotetext{a}{We include the FWHM of the objects in the continuum image
by request, but bear in mind that the beam FWHM is 
34\protect\arcsec$\times$29\protect\arcsec and so most of these objects
are unresolved.}
\end{deluxetable}

\clearpage

\begin{deluxetable}{lrrlrrrrr}
\tablecaption{Position Angles and Inclinations. \label{tabpa}}
\tablewidth{0pt}
\tablehead{
\colhead{} & \colhead{$a$\tablenotemark{a}} & \colhead{}
& \colhead{PA} 
& \colhead{} & \colhead{i\tablenotemark{b}} & \colhead{Center}
& \colhead{$\Delta$RA}
& \colhead{$\Delta$DEC}
\nl
\colhead{Component} 
& \colhead{(\protect\arcmin)}
& \colhead{$a$/R$_{25}$}
& \colhead{(\arcdeg)} 
& \colhead{$b/a$} & \colhead{(\arcdeg)} & \colhead{(1950)}
& \colhead{(\protect\arcsec\ E)}
& \colhead{(\protect\arcsec\ N)}
}
\startdata
Stars--outer isophote & 4.3 & 1.64 & 32.5 & 0.42 & 72
                                  & 7$^h$ 23$^m$ 30.9$^s$  69\arcdeg\ 18\arcmin\ 30\arcsec & 0\xx & 0\xx \nl
Stars--middle isophote & 1.6 & 0.63 & 30   & 0.27 & \nodata 
                                  & 7$^h$ 23$^m$ 34.5$^s$  69\arcdeg\ 18\arcmin\ 38\arcsec & $+$19\xx & $+$8\xx \nl
Stars--inner isophote & 1.0 & 0.37 & 32   & 0.34 & \nodata 
                                  & 7$^h$ 23$^m$ 37.2$^s$  69\arcdeg\ 19\arcmin\ 01\arcsec & $+$33\xx & $+$31\xx \nl
HI--morphology     & 7.9 & 3.02 & 23.5 & 0.41 & 73      
                                  & 7$^h$ 23$^m$ 39.8$^s$  69\arcdeg\ 19\arcmin\ 24\arcsec & $+$47\xx & $+$54\xx \nl
HI--kinematics     & $>$2.5 & $>$0.96 & 46   & \nodata & 65
                                  & 7$^h$ 23$^m$ 31.8$^s$  69\arcdeg\ 18\arcmin\ 47\arcsec & $+$5\xx & $+$17\xx \nl
\enddata
\tablenotetext{a}{Semi-major axis of the bar or, for the HI kinematics,
radius at which the measurements apply.}
\tablenotetext{b}{Calculated assuming an intrinsic $b/a$ of 0.3.}
\end{deluxetable}

\clearpage

\begin{deluxetable}{lc}
\tablecaption{Summary of Global Optical, Near-IR, and HI Properties. 
\label{tabglobal}}
\tablewidth{0pt}
\tablehead{
\colhead{Parameter} & \colhead{Value} 
}
\startdata
D (Mpc) \dotfill                                                               &  3.44 \nl
M$_{HI}$ (M\protect\solar) \dotfill                                            &  8.0$\times10^8$ \nl
E(B$-$V)$_f$\tablenotemark{a} \dotfill                                                          &  0.043 \nl
R$_{25}$\tablenotemark{b} (arcmin) \dotfill                                                 &  2.6 \nl
R$_H$\tablenotemark{b} (arcmin) \dotfill                                        &  5.1 \nl
$\mu_0$ (V-band, magnitudes arcsec$^{-2}$) \dotfill                          & 22.82$\pm$0.09 \nl
R$_D$ (kpc) \dotfill                                                           &  1.59$\pm$0.04 \nl
M$_{B,0}$ (R$=$7.4\protect\arcmin) \dotfill                                    & $-$16.51 \nl
(U$-$B)$_0$ (R$=$7.4\protect\arcmin) \dotfill                                  & $-$0.40 \nl
(B$-$V)$_0$ (R$=$7.4\protect\arcmin) \dotfill                                  & 0.31 \nl
M$_{J,0}$ (R$=$4.0\protect\arcmin) \dotfill                                    & $-$17.53 \nl
(J$-$H)$_0$ (R$=$2.6\protect\arcmin) \dotfill                                  & 0.43 \nl
(H$-$K)$_0$ (R$=$1.9\protect\arcmin) \dotfill                                  & 0.04 \nl
log L$_{H\alpha,0}$ (ergs s$^{-1}$) \dotfill                                   & 40.27 \nl
SFR\tablenotemark{c}~ (M\protect\solar\ yr$^{-1}$) \dotfill                     & 0.13 \nl
log SFR/area\tablenotemark{c}~ (M\protect\solar\ yr$^{-1}$ kpc$^{-2}$) \dotfill & $-$2.22 \nl
M$_{HI}$/L$_B$ (M\protect\solar/L\protect\solar) \dotfill                      & 1.36 \nl
L$_{H(-0.5)}$\tablenotemark{d} (L\protect\solar) \dotfill                      & 6$\times10^7$ \nl
\protect\mhilh\tablenotemark{d}~ (M\protect\solar/L\protect\solar) \dotfill     & 13 \nl
\enddata
\tablenotetext{a}{E(B$-$V)$_f$ is foreground reddening due to the Milky Way
(Burstein \& Heiles 1984).
For the stars in NGC 2366, we assume a an additional internal reddening of 0.05
magnitude;
for the \protect\HII\ regions
we assume an additional internal reddening 0.1 magnitude (consistent with 
Balmer decrement observations of the emission nebula NGC 2363; 
see references in Table 7 of Hunter \& Hoffman [2000]).}
\tablenotetext{b}{At NGC 2366 1\protect\arcmin\ $=$ 1.0 kpc.}
\tablenotetext{c}{Star formation rate derived from L$_{H\alpha}$
using the formula of Hunter \& Gallagher (1986) that integrates
from 0.1 M\protect\solar\ to 100 M\protect\solar\ with a Salpeter
(1955) stellar initial mass function.
The area is $\pi$R$_{25}^2$.}
\tablenotetext{d}{L$_{H(-0.5)}$ is measured within 0.316R$_{25}$, and 
M$_{H_{\mathord\odot}}$$=$3.44.}
\end{deluxetable}

%
%
%

\clearpage

\figcaption{Integrated velocity profile made by summing flux over
square boxes 10\protect\arcmin\ and 30\protect\arcmin\ on a side.
The larger box is approximately the primary beam size of the VLA telescope.
The fluxes are corrected for the attenuation by the primary beam.
This is compared to the profiles obtained with a single-dish radio
telescope with a 10\protect\arcmin\ beam by Hunter, Gallagher, \& Rautenkranz (1982).
\label{figsingle}}

\figcaption{21-cm continuum map shows the continuum sources within
the primary beam of the VLA data.
The map has been corrected for the attenuation by the primary beam.
The continuum sources are numbered as given in Table \protect\ref{tabcont}.
The beam size FWHM (33.6\protect\arcsec$\times$28.9\protect\arcsec)
is shown as the tiny elliptical contour in the lower left corner
near 7$^h$ 26$^m$ 20$^s$, 69\protect\arcdeg, 01\protect\arcmin.
\label{figcontfull}}

\figcaption{H$\alpha$ image of NGC 2366 shown superposed with contours
from the 21 cm continuum map. 
Contours are 1 to 7 mJy beam$^{-1}$ in steps of 1 mJy beam$^{-1}$.
The contour in the lower left corner represents the FWHM of the synthesized
beam 
(33.6\protect\arcsec$\times$28.9\protect\arcsec). 
\label{figcont}}

\figcaption{V-band image of NGC 2366. Contours are of the V-band image smoothed
14$\times$14 pixels. The inner contour is of a surface brightness level of 25.66 
magnitudes
arcsec$^{-2}$, and the outer is 26.54 magnitudes arcsec$^{-2}$.
The inner contour highlights the region fit as a bar, and
the smooth ellipse is a bar structure with $c=3.0$ and a semi-major axis
of 4.3\protect\arcmin.
The outer contour highlights the region of low surface brightness that
extends the major axis to the southwest.
For orientation, one can use the upper of two tiny elliptical contours
within the body of the galaxy to note the approximate location of NGC 2363.
Residuals of a saturated star are visible along the top edge 
of the image.
\label{figouterbar}}

\figcaption{V-band image of NGC 2366. Contours are of the V-band image smoothed
14$\times$14 pixels, at surface brightness levels of 
25.66, 24.04, 23.04, and 22.68 magnitudes arcsec$^{-2}$.
The smooth ellipses are the fits to bar structures discussed in the text
with $c=3.0$.
\label{figallbars}}

\figcaption{Surface brightness cuts along the inner bar in the V-band image of 
NGC 2366 (Figure \protect\ref{figallbars}).
Cuts along the major axis (position angle 32.5\arcdeg)
and along the minor axis are shown.
Both were constructed by summing over 33\protect\arcsec.
The spike in the major-axis profile is due to the supergiant \protect\HII\
complex NGC 2363.
The left side of the graph refers to the southeast side of the minor axis
and the northeast side of the major axis.
The vertical dashed lines mark the semi-major axis of the inner bar.
\label{figbarcut}}

\figcaption{UBV surface photometry of NGC 2366.
The photometry is corrected for an internal reddening of 0.05 mag
and an external reddening of 0.043 mag.
The exponential fit to
the V-band surface brightness profile is shown as a solid line.
The spike at 1.2\arcmin\ is due to the supergiant \protect\HII\ complex
NGC 2363.
\label{figubv}}

\figcaption{JHK surface photometry of NGC 2366. The fit to the 
V-band surface brightness
profile is shown for comparison; it has been
shifted vertically to match the J-band brightness.
The spike at 1\arcmin\ is due to the supergiant \protect\HII\ complex
NGC 2363.
\label{figjhk}}

\figcaption{False-color display of the optical B/V ratio image,
averaged $3\times3$ pixels to improve signal-to-noise.
Blue color in the display denotes a higher B/V (bluer) ratio in the galaxy.
The exceptions are regions of bright foreground stars
that produce the large
red circular region at $\sim$7$^h$ 23$^m$ 30$^s$, 69\arcdeg\ 21\arcmin\
and
the mess along the upper edge to right of middle of the display.
Contours are from the V-band image smoothed 14$\times14$ pixels and are 
26.54, 25.66, 24.04, 23.04, and 22.68 magnitudes arcsec$^{-2}$;
these are the same contours as shown in 
Figures \protect\ref{figouterbar} and \protect\ref{figallbars}.
\label{figdivbv}}

\figcaption{Azimuthally-averaged H$\alpha$ surface brightness and 
reddening-corrected V-band
surface brightness. The H$\alpha$ luminosity is proportional to the
star formation rate and was corrected for extinction using an A$_{H\alpha}$
of 0.36 magnitude.
The scales for $\Sigma_{H\alpha,0}$ and $\mu_V^0$ have been set
so that they cover the same logarithmic interval.
The spike at 1.2\arcmin\ is due to the supergiant \protect\HII\ complex
NGC 2363.
\label{figha}}

\figcaption{Channel maps of HI line-emission in NGC 2366 made
from observations with the VLA C and D-arrays, centered on the
optical galaxy. The inner 19\protect\arcmin$\times$19\protect\arcmin\ are shown.
The beam size FWHM
(33.6\protect\arcsec$\times$28.9\protect\arcsec)
is shown in the final panel of the channel maps.
\label{figchan}}

\figcaption{Integrated HI map of NGC 2366.
Shallow display (left) shows the central HI ring; 
deep display (right) shows structure in the outer HI.
The contour is of the V-band image of the galaxy smoothed
14$\times$14 and at a surface brightness level of 25.66 magnitudes
arcsec$^{-2}$.
The FWHM of the synthesized beam, shown as the ellipse in the lower
left corner of each panel, is
33.6\arcsec$\times$28.9\arcsec.
\label{figgrayfull}}

\figcaption{Surface density of HI for NGC 2366 is computed in two
ways: 
closed circles---using the position angle (23\arcdeg) and inclination
(73\arcdeg) of the HI morphology;
open circles---using the
position angle (46\arcdeg) and inclination (65\arcdeg) of the
HI kinematics. Left:
For comparison we show the average HI surface density profile of  a sample
of Im, Sdm, and Sd galaxies. Those data are from Cayatte \protect\et\ (1994) 
and Broeils \& van Woerden (1994).
These references normalize the radius to
R$_o$, a radius given by de Vaucouleurs \et\ (1991).
In most cases R$_o$ is nearly the same as R$_{25}$, the
normalization we use for NGC 2366.
The x-axis at the top of the figure is R in arcminutes for NGC 2366 only.
Right: In order to compare the HI surface density with that of H$\alpha$
and the starlight, we show the logarithm of $\Sigma_{HI}$.
The V-band and H$\alpha$ surface brightness profiles of NGC 2366 are also shown. 
The scale for the H$\alpha$ surface brightness, not shown, goes from
a logarithm of 30.75 to 34.25 with units of ergs s$^{-1}$ pc$^{-2}$.
The plotting scales are chosen so that a magnitude interval corresponds to
0.4 of the invertal for logarithmic quantities and all logarithmic
quantities are plotted over the same number of dex.
\label{figsurden}}

\figcaption{HI column density contours superposed on our H$\alpha$ image.
The HI contours are 0.88, 3.8, 6.8, 9.7, 15.6, 21.5, 27.4, 33.3, 39.2,
and 45.0 $\times10^{20}$ atoms cm$^{-2}$. 
The ellipse in the lower left corner is the FWHM of the synthesized
beam (33.6\protect\arcsec$\times$28.9\protect\arcsec).
\label{fighionha}}

\figcaption{Brightnesses of individual pixels are compared in HI and
\protect\ha.
The \ha\ image was
averaged 15$\times$15 to produce a pixel scale of 6.3\protect\arcsec,
close to the pixel scale of the HI map.
The x-axis is the HI pixel brightness in mJy beam$^{-1}$ where
1 mJy beam$^{-1}$ corresponds to
a column density of 5.9$\times10^{18}$ \protect\coldens.
The y-axis is the logarithm of the 
\protect\ha\ pixel brightness in counts
in the image
where 1 count corresponds to an \protect\ha\ luminosity of
1.15$\times10^{33}$ ergs s$^{-1}$, corrected for reddening.
\label{figimvim}}

\figcaption{Color-coded display of the HI velocity
field.
The beam size (FWHM shown in the lower left corner) is
33.6\arcsec$\times$28.9\arcsec\
and the velocity resolution is 10 \protect\kms.
In the upper panel
the contours are of the model best-fit velocity field at 60, 80, 100, 
120, and 140 \protect\kms.
In the bottom panel, the outer contour is the inner V-band contour shown 
in Figure \protect\ref{figouterbar}; and the
inner contour is of the \protect\ha\ image.
The units of the color wedge are \protect\kms.
\label{figvel}}

\figcaption{Top panel: Best-fit rotation curve. The rotation curve
comes from fitting rings of 20\protect\arcsec\ width out to a radius of
460\protect\arcsec\ and 40\protect\arcsec\ width beyond that to a radius
of 620\protect\arcsec.  The center position, systemic velocity (99
\protect\kms), position angle (46\arcdeg), and inclination (65\arcdeg)
were fixed in the final fit. The kinematics of the approaching side of
the galaxy are peculiar and not well fit by the model.  
The horizontal dashed line at 40 \protect\kms\ is an aid to the eye
in following the rotation curve.
Middle panel:
Position angle for tilted ring model fits to the velocity field.
A position angle was determined from the central part of the velocity
field with emphasis on the well-behaved receding part of the galaxy.
The dashed horizontal line marks the value of the final position angle.
Bottom panel: Inclination for tilted ring model fits to the velocity
field.  An inclination was determined from the central part of the
velocity field with emphasis on the well-behaved receding part of
the galaxy.  The dashed horizontal line marks the value of the final
inclination angle.  \label{figrot}}

\figcaption{Color-coded display of the HI velocity dispersion.
The small ellipse in the lower, left corner is the FWHM of the beam
(33.6\arcsec$\times$28.9\arcsec).
In the top panel the outer contour is the outer V-band contour shown in Figure
\protect\ref{figouterbar}. The inner contour is of the \protect\ha\ image
and corresponds to 
1.7$\times10^{34}$ ergs s$^{-1}$ arcsec$^{-2}$.
In the middle panel the contours are of the velocity field map and
correspond to 60 \protect\kms\ to 140 \protect\kms\ in steps of
10 \protect\kms.
In the bottom panel the contours outline the integrated HI map and are
3.5, 7.1, 10.6, 14.1, 21.5, 33.3, and 45.0$\times10^{20}$ \protect\coldens.
The units of the color wedge are \protect\kms.
\label{figvdispha}}

\figcaption{HI line profiles for the region of high HI velocity
dispersion on the eastern edge of the HI distribution.
Every five pixels, approximately one beam-width, have been averaged.
Flat-top profiles are profiles whose tops have been chopped off 
in plotting in order to be able to show the weaker profiles.
\label{figplcubeast}}

\figcaption{East--west scans of HI column densities (dashed lines) and velocity
dispersions (solid lines) across the disk of NGC 2366, averaged over
one beam width in declination. Each goes through a region where
the velocity dispersion has a sudden increase in the outer part of the
disk. In the main part of the galaxy, the 
velocity dispersion anticorrelates with the column density,
suggesting that turbulent motions are faster in the lower density regions.
\label{vdcolumn}}

\figcaption{Region of the supergiant \protect\HII\ complex
NGC 2363. 
The gray-scale is H$\alpha$, and the contours are 
the integrated C-array HI.
The contours are 2.43$\times10^{21}$ \protect\coldens\
to 6.68$\times10^{21}$ \protect\coldens, in steps of
6.075$\times10^{20}$ \protect\coldens.
The ellipse in the lower left corner is the FWHM of the HI beam,
17.8\arcsec$\times$13.2\arcsec\ (300 pc$\times$220 pc at the galaxy).
\label{fign2363}}

\figcaption{Region of the galaxy including the supergiant
\protect\HII\ complex NGC 2363.
The gray-scale is H$\alpha$, and the contours are 
the C-array HI velocity field. The contours
begin with 60 \kms\ at the far right and go to 105 \kms\ in the upper
left corner in steps of 5 \kms.
The ellipse in the lower right corner is the FWHM of the HI beam,
17.8\arcsec$\times$13.2\arcsec\ (300 pc$\times$220 pc at the galaxy).
\label{fig2363vel}}


\figcaption{HI line profiles from the C-array data
for a region centered on NGC 2363. The data cube
has been averaged over five pixels (15\protect\arcsec),
approximately one beam-width.
\label{fign2363plcub}}

\figcaption{Surface densities in NGC 2366. 
The dashed vertical lines mark radii where the rotation curve
(Figure \protect\ref{figrot})
ceases to be that of a solid body (left) and where it begins
to turn over (right).
The critical gas density
for the thin-rotating disk model \sigcrit\ is determined from the 
rotation curve of the galaxy and the mean velocity dispersion, determined
as a function of radius from the C$+$D-array second moment map.
The azimuthally-averaged
gas surface density \protect\siggas\ is determined assuming a position angle of
46 \protect\arcdeg\ and an inclination of 65\protect\arcdeg\ and is
corrected for the presence of He and, statistically, for H$_2$.
The \protect\sighii\ are the maximum surface densities of the gas
in the vicinity of each \protect\HII\ region in the galaxy.
The \protect\sigha\ are the H$\alpha$ surface brightnesses measured
from the H$\alpha$ image using the optically-derived position angle
of 32.5\protect\arcdeg\ and 72 \protect\arcdeg. \protect\sigha\
shows where star formation is taking place within the galaxy.
The $\Sigma_T$ are limiting tidal densities.
\label{figsig}}

\end{document}